\documentclass[%
 reprint,
 amsmath,amssymb,
 aps,
]{revtex4-2}

\usepackage{graphicx}
\usepackage{dcolumn}
\usepackage{bm}
\usepackage{subfigure}
\usepackage{caption}
\usepackage{subcaption}
\usepackage{float}
\usepackage[utf8x]{inputenc}
\usepackage{tikz}
\usepackage{xcolor}
\usepackage{amsmath}
\usepackage{pgfplots}
\pgfplotsset{compat=1.18}
\usetikzlibrary{arrows.meta, decorations.pathmorphing, positioning}

\begin{document}

\preprint{APS/123-QED}

\title{Deflection of Light due to Kerr Sen Black Hole in Heterotic String Theory\\ using Material Medium Approach}

\author{Saswati Roy}
  \email{sr.phy2011@yahoo.com}
 \affiliation{Department of Physics, National Institute of Technology, Agartala, Tripura,799046, India}
 
\author{Shubham Kala}%
 \email{shubhamkala871@gmail.com}
\affiliation{The Institute of Mathematical Sciences, C.I.T. Campus, Taramani, Chennai, 600113, India}

\author{Atanu Singha}
\email{atanusingha008@gmail.com}
\affiliation{
 Department of Physics, National Institute of Technology, Agartala, Tripura,799046, India}%

\author{Hemwati Nandan}
    \email{hnandan@associates.iucaa.in}
	\affiliation{Department of Physics, Hemvati Nandan Bahuguna Garhwal Central University, Srinagar Garhwal, Uttarakhand 246174, India}
		\affiliation{Center for Space Research, North-West University, Potchefstroom 2520, South Africa}

\author{A. K. Sen}
    \email{asokesen@yahoo.com}
	\affiliation{Department of Physics, Assam University, Silchar, Assam 788011, India}

\date{\today}

\begin{abstract}
The deflection of light in the gravitational field of a massive body can be analyzed through diverse theoretical approaches. The null geodesic approach is commonly employed to calculate light deflection within strong and weak field limits. Alternatively, several studies have explored the gravitational deflection of light using the material medium approach. For a static, non-rotating spherical mass, the deflection in a Schwarzschild field can be determined by expressing the metric in an isotropic form and evaluating the refractive index to trace the light ray's trajectory. In this study, we extend the above-mentioned approach to the Kerr-Sen black hole spacetime in heterotic string theory, a solution representing a rotating, charged solution in heterotic string theory. The frame-dragging effects inherent to the Kerr-Sen geometry are incorporated to compute the velocity of light rays, enabling the derivation of the refractive index in this field. Considering the far-field approximation, we calculate the deflection of light in the Kerr-Sen spacetime and compare our results with those obtained for the Kerr and Schwarzschild black hole solution in GR.  
\end{abstract}


\maketitle

\section{Introduction}

The phenomenon of bending of light when it passes around massive objects due to gravity is one of the important predictions of General Relativity (GR). Gravitational Lensing (GL) is the term used to describe the phenomenon that arises when electromagnetic radiation in a gravitational field is deflected, and a gravitational lens is an object that causes a discernible deflection. Since the GL is independent of the nature and physical condition of the lensing mass, it offers a clear and distinct way to inquire the presence of dark matter at all distance scales, making it one of the most significant fields in modern astronomy.\\
Another fascinating prediction of GR is the existence of super massive highly compact object known as black hole (BH).  The various BH solutions in GR include the static neutral case of Schwarzschild BH (SBH), which was the first precise solution to the GR field equations of Einstein of a single spherical non-rotating mass given by Schwarzschild \cite{schwarzschild1916gravitationsfeld}. Later, different solutions were also obtained, such as the neutral and rotating solution\cite{kerr1963gravitational}, the non-rotating charged solution\cite{reissner1916eigengravitation, nordstrom1918energy, janis1968reality} and finally the charged rotating BH solution\cite{newman1965metric}. In addition to BH in GR, there are others that arise from several alternate theories of gravity such as scalar-tensor theory \cite{fujii2003scalar}, string theory \cite{tong2009lectures}, braneworld scenario \cite{sengupta2008aspects}, and loop quantum gravity \cite{thiemann2003lectures}. Specifically, the majority of BHs described by one or more charges connected to Yang–Mills fields are emerging in string theory \cite{mignemi1993charged, gibbons1988black}, which unifies gravity with the other three fundamental forces in nature. This means that compared to the GR, such stringy BHs may offer far deeper insight into the true nature of gravity. The Kerr-Sen BH (KSBH) \cite{sen1992rotating}, known as the rotating charged BH solution in the low-energy limit of string field theory, is a dilaton-axion generalization of the well-known Kerr BH (KBH) in GR.\\
In recent years, various studies have been conducted to understand the KSBH spacetime in diverse contexts. Bhadra \cite{Bhadra:2003zs} investigated lensing due to the charged string black hole, known as  Gibbons–Maeda–Garfinkle–Horowitz–Strominger black hole (GMGHSBH) and also mentioned how it differs with the Reissner-Nordstr\"om black hole (RNBH).
Gyulchev and Yazadjiev \cite{gyulchev2007kerr, gyulchev2010analytical} have studied quasi-equitorial lensing by KSBH in the strong and weak deflection limit, furthermore, they compared with Schwarzschild, Kerr and Gibbons-Maeda BHs. In 2015, Siahaan \cite{siahaan2016destroying} discussed the possibility of destroying a near extremal KSBH, i.e. applying Wald's \cite{wald1974gedanken} gedanken experiment, they have shown that a near extremal KSBH can turn into a naked singularity by capturing spinning and massive charged particles. Xavier et al. \cite{xavier2020shadows} conducted a detailed study of the shadows of the charged rotating BH solution from GR i.e. Kerr-Newman BH (KNBH) solution in contrast to the KSBH solution and concluded that the latter has a larger shadow for similar physical parameters and observation conditions. Also in 2020, Minyong Guo et al. \cite{guo2020observational}, for the case of a near extremal KSBH, studied the appearance of an isotropically emitter orbiting close to its horizon and the effects of Kerr-Sen charge or string charge on quantities such as the redshift factor and image position and further compared their results with a near extremal KNBH. Recently, a few authors have undertaken the study of gravitational deflection of light around a KSBH.  Uniyal et al. performed \cite{uniyal2017null, uniyal2018bending} a thorough analysis of GL by a KSBH in the equatorial plane and derive an exact closed-form solution for the light deflection angle that satisfies the weak and strong field limits. They have succeeded in formulating the bending angle explicitly in both direct and retrograde motion scenarios, where the BH mass and spin parameters determine the final bending angle expression by using the null-geodesic method.\\
The deflection of the light ray can be calculated using different methods. The precise deflection of a light beam, when it approaches a gravitational mass, is generally obtained by using the null geodesic through which the beam travels \cite{misner1973gravitation, weinberg1972principles, schneider1992gravitational}. An alternative method to calculate the deflection of a light beam near a gravitational mass is to assume that the light beam is traveling through a material medium, the refractive index of which is determined by the gravitational field. This method estimates the impact of gravitation on the light beam. This is known as the \textit{Material Medium Approach}.\\
The idea of material medium approach was first introduced in the year 1924 \cite{tamm1924electrodynamics}. 
Later on, this idea was discussed by different authors in order to determine how a spinning body affects the polarization of an electromagnetic wave and investigate how a gravitational field scatters a plane electromagnetic wave \cite{balazs1958effect, plebanski1960electromagnetic}, to compute the polarization and deflection caused by the SBH and KBHs \cite{mashhoon1975influence, fischbach1980second}, and computed the gravitational time delay and light ray paths in SBH geometry using the effective refractive index \cite{evans1986f, evans1996optical}.
Using the optomechanical analogy of GR, the Newtonian formalism of stationary metrics was expanded, particularly of spinning spacetimes \cite{evans1986f}. Faraday rotation and GL were also discussed in the weak field limit using Fermat’s principle \cite{sereno2003gravitational, sereno2004weak}.
In 2010, Sen \cite{sen2010more} obtained the light deflection angle for a static non-rotating mass in SBH geometry using the material medium approach, without the need for any approximation in terms of elliptic integral. In line with this, Roy and Sen computed an analytical expression of the refractive index in terms of the rotation parameter and hence the deflection angle for a rotating mass in the KBH geometry \cite{roy2015trajectory} and for the RN and Janis-Newman-Winicour (JNW) spacetime geometries \cite{roy2017deflection}. Very recently also, this method was used to analyze the gravitational deflection of light rays and Shapiro time delay \cite{del2024accurate} and to study the interaction of a plane gravitational wave with electromagnetic fields \cite{ruggiero2025effects}. \\
\begin{figure}[H]
    \centering
    \begin{subfigure}[]
        \centering
        \begin{tikzpicture}[scale=0.7]

        \def\rBH{1.0}
        \def\rVac{2}
        \def\rMed{3}

        \fill[blue!15] (0,0) circle (\rMed);         
        \fill[gray!30] (0,0) circle (\rVac);         
        \fill[white] (0,0) circle (\rBH);            
        \fill[black] (0,0) circle (\rBH);            

        \draw[very thick, red, -{Stealth}] 
           (-3.5,1.2) 
           to[out=5, in=160] 
           (0.3,1.0) 
           to[out=-20, in=150] 
           (3.5,-0.5);

        \node at (0.6,0.3) [below left, text=white, font=\small\bfseries] {BH};
        \node[gray!50!black] at (0.1, -1.3) {\small Vacuum Region};
        \node[blue!60!black] at (3.8, 0.8) {\small Material Medium};
        \node[red] at (-2.2,1) [left] {\small Incident Light Ray};
        \node[red] at (2.0,-0.7) [right] {\small Deflected Light Ray};

        \draw[dashed] (0,0) circle (\rVac);
        \draw[dashed] (0,0) circle (\rMed);

        \draw[->, thick, blue!60!black] (1.5, 2.5) -- (1.0, 1.7);
        \node[blue!60!black] at (2.3, 2.5) {\small Graded Refractive Index $n(r)$};

        \end{tikzpicture}
        \label{fig:schematic}
    \end{subfigure}
    \begin{subfigure}[]
        \centering
        \begin{tikzpicture}
        \begin{axis}[
            width=6cm,
            height=5.5cm,
            axis lines=left,
            xlabel={$r$},
            ylabel={$n(r)$},
            ymin=0.9, ymax=2.5,
            xmin=1, xmax=6,
            samples=100,
            domain=1.01:6,
            smooth,
            thick,
            every axis plot/.append style={blue!70!black},
            label style={font=\small},
            tick label style={font=\small},
            minor tick num=1,
        ]
        \addplot[blue, thick] {1 + 1/(x^2)};
        \end{axis}
        \end{tikzpicture}
    \end{subfigure}
     \caption{(a) Schematic view for the deflection of light due to a graded refractive index near a BH. Here, light bends because it follows the path of least optical distance in the refractive medium and vacuum acts as an effective medium influenced by the gravitational field.(b) The sample refractive index profile indicating the decline of refractive index, n(r) with the distance.}
    \label{schematic}
\end{figure}
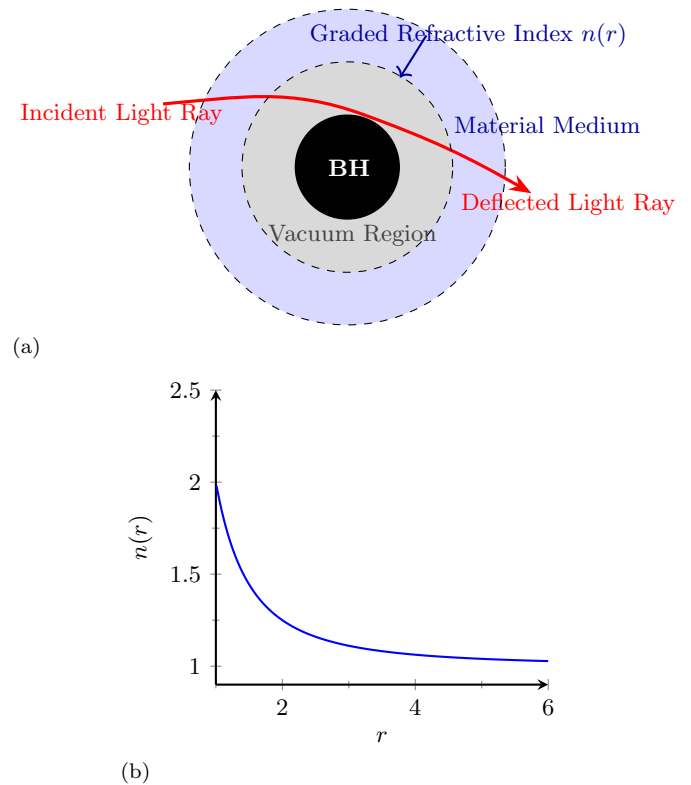
The organization of the paper is as follows. Section 2 provides an overview and discussion of the KSBH spacetime, laying the groundwork for our analysis. In Section 3, we numerically solve for the photon sphere and present its graphical interpretation. Section 4 focuses on the calculation of the velocity of light in the KSBH field and the determination of the associated refractive index. In Section 5, we calculate the deflection angle of light influenced by the KSBH. Finally, in Section 6, we summarize the results and conclude the paper.
\section{Kerr Sen Black Hole}
The KSBH is a four-dimensional solution of KBH arising from the heterotic string theory, which describes a rotating, electrically charged massive entity. In this view, the appropriate effective action\cite{schneider1992gravitational} is as follows,
\begin{equation} \label{Eq:1}
    S =\int_{}^{}d^4x \sqrt{|\Tilde{g}|}e^{-\Tilde{\Phi}}(R-\frac{1}{12}H^2 +\Tilde{g}^{\mu\nu}\partial_\mu\Tilde{\Phi}\partial_\nu\Tilde{\Phi}- \frac{1}{8}F^2),
\end{equation}
where $g$ is the determinant of the metric tensor $g_{\mu\nu}$, $F^2$ is the square of field-strength tensor $F_{\mu\nu}=\partial_\mu +A_\nu$, $\Phi$ is the dilaton; $H^2$
is the square of a third rank tensor field
\begin{equation} \label{Eq:2}
H_{\kappa\mu\nu}=\partial_\kappa B_{\mu\nu} + \partial_\nu B_{\kappa\nu} + \partial_\mu B_{\nu\kappa} -\frac{1}{4}(A_\kappa F_{\mu\nu} + A_\mu F_{\nu\kappa} + A_\nu F_{\kappa \mu}),
\end{equation}
where $B_{\mu\nu}$ is a second rank anti-symmetric tensor field. It is evident that the Einstein-Hilbert action results when all non-gravitational fields in the action (\ref{Eq:1}) vanish. The Kerr-Sen metric is a solution in the theory given by (\ref{Eq:1}) when all non-gravitational fields are absent since it satisfies the vacuum Einstein equations. In fact, by implementing a series of transformations that link the solution in (\ref{Eq:1}) to the KBH metric, the KSBH solution is produced \cite{siahaan2016destroying}.
 
The KSBH metric provides the precise solution to Einstein's Field Equation of GR for a stationary, axially symmetric gravitational field of a rotating body in a dilaton field. In the Boyer-Lindquist coordinates ($ct, r, \theta, \phi$), the KSBH line element can be represented as follows,
\begin{equation} \label{Eq:3}
\begin{split}
    ds^2 = &\frac{\Delta-\alpha^2 \sin^2{\theta}}{\Sigma^2}c^2dt^2 - \frac{\Sigma^2}{\Delta}dr^2 - \Sigma^2d\theta^2 \\
    &-\left(\Delta +\frac{r_gr(r(r+2b)+\alpha^2)} {\Sigma^2}\right) \sin^2{\theta} d\phi^2 \\&+ \frac{2r_gr\alpha\sin^2{\theta}}{\Sigma^2}cd\phi dt, 
\end{split}
\end{equation}
where the metric functions are described as follows,
\\
\begin{equation} \label{Eq:4}
\begin{split}
    &\Sigma^2 =r(r+2b)+\alpha^2\cos^2{\theta}, \\
    \\
    &\Delta =r(r+2b)-r_gr+\alpha^2, \\
    \\
    &b =\frac{Q^2}{2m}=\frac{Q^2}{r_g}, \\
    \\
    &\alpha =\frac{J}{Mc}, \\
    \\
    &r_g =\frac{2GM}{c^2}. 
\end{split}
\end{equation}
The electric charge of the BH is represented by the symbol $Q$ (in the dimension of length), BH's rotational parameter $\alpha$ (in the dimension of length) is expressed as the ratio of its angular momentum ($J$) to its  mass ($M$) and the constant $r_g=2m$ is known as the schwarzschild radius. The coefficients of the line element (\ref{Eq:3}) exhibit independence from $\phi$, indicating its axial symmetry. The spacetime characterized by the KSBH metric is not vacuum, similar to the KNBH situation in the Einstein-Maxwell theory.  In the theory represented by the action (\ref{Eq:1}), the solutions for non-gravitational basic fields are as follows \cite{siahaan2016destroying},

\begin{equation} \label{Eq:5}
    \Tilde{\Phi}=-\frac{1}{2}\ln{\frac{\Sigma^2}{r^2+\alpha^2\cos^2{\theta}}},
\end{equation}

\begin{equation} \label{Eq:6}
    A_t=\frac{-rQ}{\Sigma^2},
\end{equation}

\begin{equation} \label{Eq:7}
    A_\phi=\frac{rQ\alpha \sin^2{\theta}}{\Sigma^2}
\end{equation}

\begin{equation} \label{Eq:8}
    B_{tQ}=\frac{br\alpha \sin^2{\theta}}{\Sigma^2}
\end{equation}
\\
Setting the parameter $b = 0$ yields the result that all non-gravitational fields (\ref{Eq:5} - \ref{Eq:8}) vanish, and the line element (\ref{Eq:3}) reduces to the Kerr metric. Furthermore, if we set the rotational parameter $\alpha = 0$, the line element (\ref{Eq:3}) is reduced to the SBH metric, which represents the electrically neutral, non-rotating, or static gravitating mass. Only by setting the rotational parameter $\alpha = 0$, the line element (\ref{Eq:3}) is reduced to the GMGHSBH, which represents the charged string black hole.

\begin{figure}[H] 
	\begin{center}       
        {\includegraphics[width=0.5\textwidth]{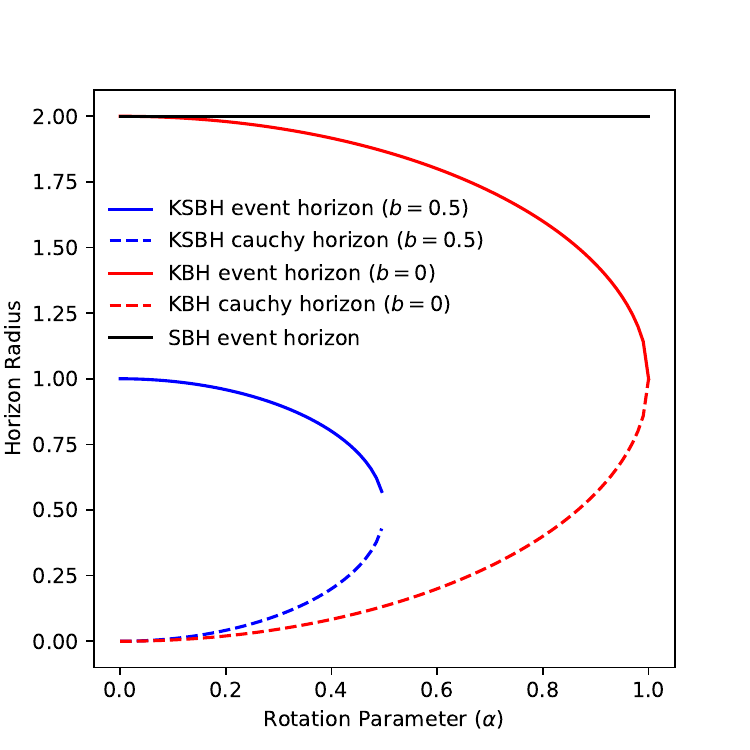}}
	\end{center}
	\caption{The variation of horizon radii with respect to the rotation parameter ($\alpha$) for KSBH as compared to other BHs horizon in GR} \label{FigHorizons}
\end{figure}

The Kerr-Sen line element  (\ref{Eq:3}) has a spherical event horizon, determined by $\Delta(r)=0$, whose outer and inner horizons are located at
\begin{equation} \label{hor_out_in}
   r_\pm = m-b\pm\sqrt{(m-b)^2-\alpha^2}.\\
\end{equation}
 Eq.(\ref{hor_out_in}) yields that $(m-b)\geq \alpha$ or $(m- \frac{Q^2}{r_g})\geq \alpha$ unless the horizon disappears. The ranges for the rotation parameter ($\alpha$) and charge ($Q$) for KSBH are $0\leq \alpha \leq m$ and $0\leq Q \leq \sqrt{2} m$.
In FIG. \ref{FigHorizons}, the horizon structure of the KSBH is illustrated and compared with other well-known BH solutions. As the rotation parameter increases, the event horizon decreases while the Cauchy horizon increases similar to the behavior observed in the KBH due to enhanced spacetime dragging. The presence of charge in the KSBH further reduces the event horizon and expands the Cauchy horizon by introducing additional repulsive effects, as compared to the KBH. As expected, the SBH, being static and uncharged, exhibits a constant event horizon, represented as a straight line in the plot. These variations in horizon structure have direct physical significance, influencing BH observables such as the deflection of light.
The KSBH differs from the KNBH in its properties, in Einstein's Gravity. Specifically, the dual axion pseudoscalar field and dilaton scalar field that distinguish the KSBH from the KNBH \cite{wu2020ultraspinning}. Furthermore, the KNBH is algebraically type-D, whereas the Kerr-Sen BH is of type-I \cite{burinskii1995some, griffiths2009exact}. The behavior of the BH's evaporation was found to be intermediate between that of the KNBH and Kaluza-Klein BH upon examination. It was seen that given identical observational and physical characteristics, the KSBH always has a wider shadow than the KNBH \cite{xavier2020shadows, zhang2021escape}.

\section{The Photon Sphere}
In this section, we study the photon sphere of the KSBH in the equatorial plane. Our analysis closely follows the approach outlined by Feng et al.\cite{feng2020size}, which utilizes linear length as a measure to investigate the relevant photon orbits. For KSBH, determining the photon sphere requires solving a quartic equation, which is expressed as
\begin{equation}
    \begin{aligned}
    4r_{ph}^{4} &+ 12 (b - 2M) r_{ph}^{3} + (36 M^{2} - 44 M b + 13 b^{2}) r_{ph}^{2} \\
    &- \left[ 16 a^{2} M - 6 b (b - 2M)^{2} \right] r_{ph}\\
    &- \left[ 8 a^{2} M - b (b - 2M)^{2} \right] b = 0.
\end{aligned}
\end{equation}
The quartic equation governing the photon sphere can yield either two real roots and two complex conjugate roots or four real roots. To ensure that the photon sphere has physical relevance, we require that it lies outside the BH event horizon, i.e., $r_{ph}>r_{+}$. Therefore, the radii of the photon sphere are determined by solving the quartic equation under this constraint as
\begin{equation}\label{PSEq}
    r_{ph}^{\pm} = \frac{3M-b}{2} + \mathcal{A} \pm \sqrt{\left( \frac{36 M^{2}-25 b^{2}-20Mb}{4} - 4 \mathcal{A}^{2} - \frac{\mathcal{B}}{\mathcal{A}} \right)}.
\end{equation}
\begin{figure}[H] 
	\begin{center}
    \begin{subfigure}[]
     {\includegraphics[width=0.4\textwidth]{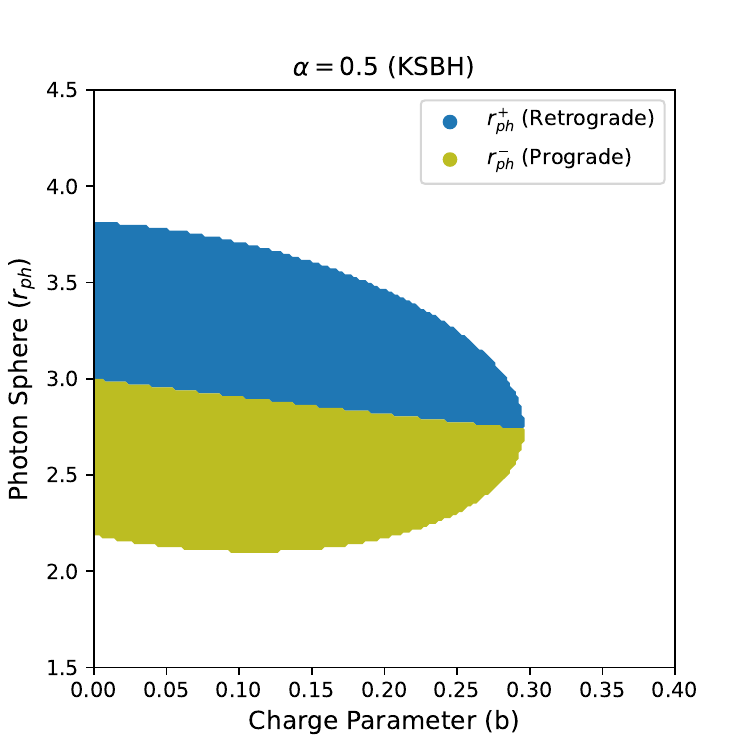}}   
    \end{subfigure}
     \begin{subfigure}[]
     {\includegraphics[width=0.4\textwidth]{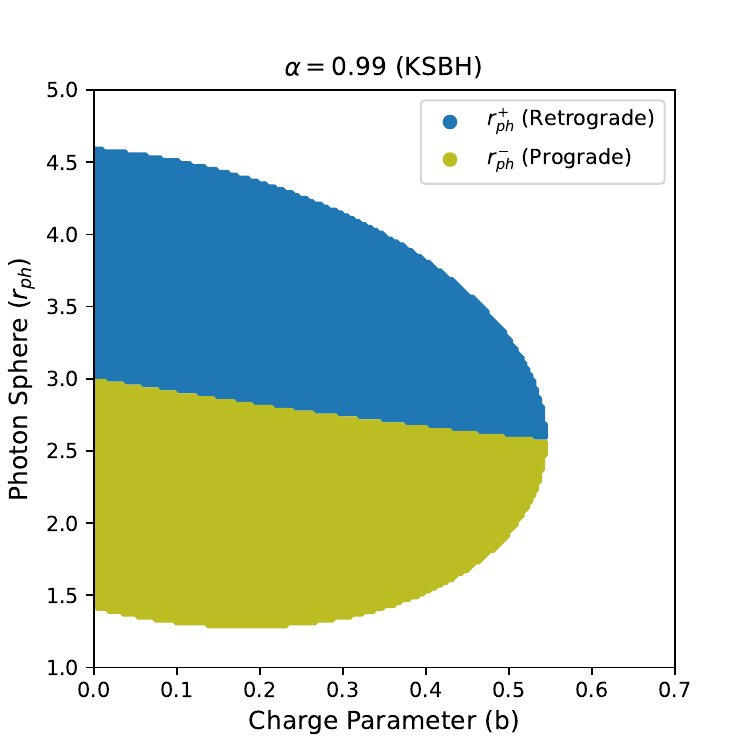}} 
    \end{subfigure}   
	\end{center}
	\caption{The possible regions of the photon sphere as a function of the charge parameter for different values of the rotation parameter.} \label{PSFig1}
\end{figure}
 The radii $r_{ph}^{+}$ and $r_{ph}^{-}$ represent the photon sphere for prograde and retrograde motion, respectively. The detailed expressions for $\mathcal{A}$ and $\mathcal{B}$ are provided in the appendix.
\begin{figure}[H] 
	\begin{center}
    \begin{subfigure}[]
     {\includegraphics[width=0.4\textwidth]{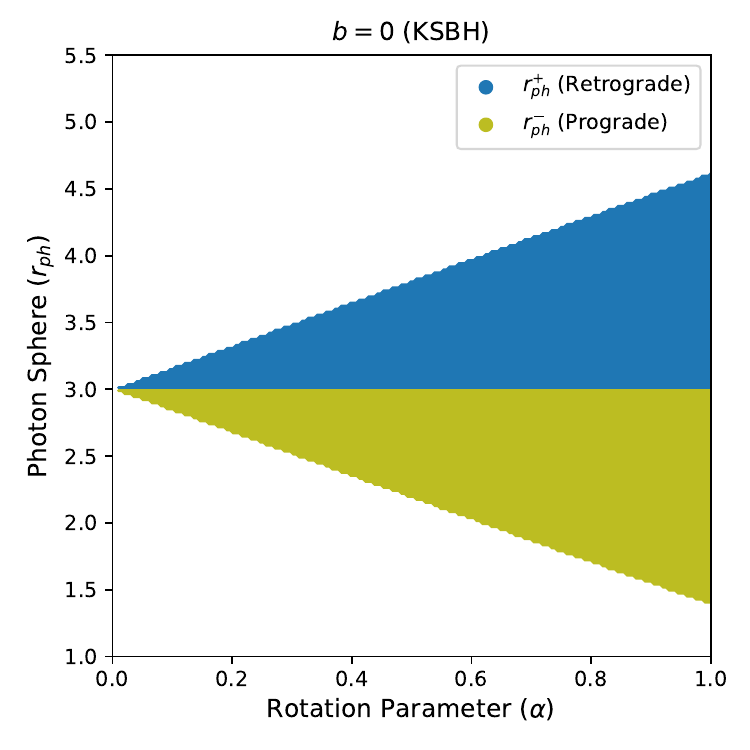}}   
    \end{subfigure}
     \begin{subfigure}[]
     {\includegraphics[width=0.4\textwidth]{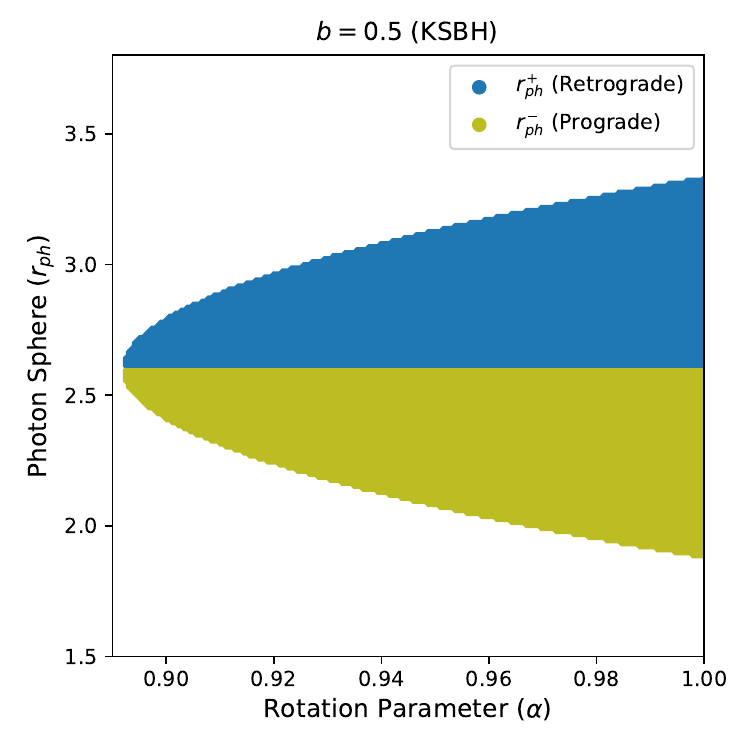}} 
    \end{subfigure}   
	\end{center}
	\caption{The possible regions of the photon sphere as a function of the rotation parameter for different values of the charge parameter.} \label{PSFig2}
\end{figure}
In the FIG. \ref{PSFig1}, we illustrate the variation of the photon sphere of the KSBH with the charge parameter. We find that the photon sphere radius decreases for the retrograde case while increasing for the prograde case as the charge parameter increases. Notably, for higher values of the rotation parameter, the retrograde photon sphere reaches its maximum, whereas the prograde photon sphere reaches its minimum. In FIG. \ref{PSFig2} we analyze the behavior of photon sphere as a function of rotation parameter. In left panel it can be easily seen that in the absence of charge parameter at $\alpha=0$, the radius of photon sphere become equal to SBH. However, when we increase the charge parameter then photon sphere of retrograde motion increased while decreased for prograde motion as we increasing the rotation parameter. In FIG. \ref{PSFig2}, we analyze the behavior of the photon sphere as a function of the rotation parameter.  In FIG. \ref{PSFig2}(a), it is evident that in the absence of the charge parameter ($b=0$), the photon sphere corresponds to that of a KBH. Additionally, for $b=0$ and $\alpha=0$, the photon sphere radius coincides with that of a SBH, confirming the standard non-rotating BH solution in GR. However, when the charge parameter is non-zero, increasing spin causes the retrograde photon sphere to expand while the prograde photon sphere shrinks. Physically, one can summarize the observation to say that the charge parameter increases the retrograde photon sphere while decreasing the prograde one, whereas the rotation parameter has the opposite effect, enhancing frame-dragging and shifting the photon sphere accordingly.
\section{Refractive Index due to Kerr-Sen BH}
The linearized form of the KSBH (\ref{Eq:3}) can be written in terms of spherical polar coordinates ($ct, r, \theta, \phi$) as
\begin{equation} 
\begin{split}
    ds^2 = &\left(\frac{\Delta-\alpha^2 \sin^2{\theta}}{\Sigma^2} + \frac{2r_gr\alpha \sin^2{\theta}}{\Sigma^2} \frac{d\phi}{cdt}\right)c^2dt^2  - \frac{\Sigma^2}{\Delta}dr^2 \\ 
    &- \Sigma^2d\theta^2  - \left(\Delta + \frac{r_gr(r(r+2b)+\alpha^2)}{\Sigma^2}\right) \sin^2{\theta} d\phi^2\\
     = & \left(1 -\frac{r_g r }{\Sigma^2} + \frac{2r_gr\alpha \sin^2{\theta}}{\Sigma^2} \frac{d\phi}{cdt}\right)c^2dt^2 - \frac{\Sigma^2}{\Delta}dr^2 - \Sigma^2d\theta^2 \\ 
    & - \left(\Delta + \frac{r_gr(r(r+2b)+\alpha^2)}{\Sigma^2}\right) \sin^2{\theta} d\phi^2.
\end{split}
\end{equation}
Under far field approximation we can assume, $\frac{\alpha^2}{r^2}<<1$ hence we get the line element as
\begin{equation} 
\begin{split}
    ds^2 = & \left(1 -\frac{r_g  }{(r+2b)} + \frac{2r_g\alpha \sin^2{\theta}}{(r+2b)} \frac{d\phi}{cdt}\right)c^2dt^2 \\ 
    & - \frac{(r+2b)}{(r+2b-r_g )}dr^2 - r(r+2b)d\theta^2 \\ 
    & - r(r+2b) \sin^2{\theta} d\phi^2.
\end{split}
\end{equation}
Again, for equatorial plane, $\theta=\frac{\pi}{2}$ which turns the line element as
\begin{equation}\label{Eq_linear} 
\begin{split}
    ds^2 = & \left(1 -\frac{r_g  }{(r+2b)} + \frac{2r_g\alpha }{(r+2b)} \frac{d\phi}{cdt}\right)c^2dt^2 \\ 
    &- \frac{(r+2b)}{(r+2b-r_g )}dr^2 - r(r+2b)\left(d\theta^2 + d\phi^2\right)\\
    = & \left(1 -\frac{r_g  }{(r+2b)} + \frac{2r_g\alpha }{(r+2b)} \frac{d\phi}{cdt}\right)c^2dt^2 \\ 
    &- \frac{1}{(1-\frac{r_g}{(r+2b)})}dr^2 - r(r+2b)\left(d\theta^2 + d\phi^2\right).
\end{split}
\end{equation}
For all intents and purposes, the gravitational field surrounding a revolving star or planet is satisfactorily described by the linearized metric. Now by coordinate transformation, one can express the above line element in isotropic form that can be used to generate the expression for refractive index as that of static field by SBH geometry\cite{sen2010more}, for a stationary field by KBH geometry \cite{roy2015trajectory} and for static charged body by RN and JNW metric \cite{roy2017deflection}.
To express the above line element, expressed by Eq. (\ref{Eq_linear}), in isotropic form, we introduce a new radial coordinate $\rho$ with the following transformation equation,
\begin{equation}\label{Eq_r}
    r= \rho(1+\frac{r_g-2b}{4\rho})^2
\end{equation}
 or,
\begin{equation} \label{Eq_rho}
    \rho= \frac{1}{2}\left[r+(b-\frac{r_g}{2})+\sqrt{(r+(b-\frac{r_g}{2}))^2-(b-\frac{r_g}{2})^2} \right]
\end{equation}
which in turn leads to, 
\begin{equation} \label{Eq_drdrho}
\begin{split}
    \frac{dr}{d\rho}&=(1+\frac{r_g-2b}{4\rho})^2-(\frac{r_g-2b}{2\rho})(1+\frac{rg-2b}{4\rho}) \\
    &=1-\frac{(r_g-2b)^2}{16\rho^2}.    
\end{split}
\end{equation}
Now, by substituting the values of $r$ and $dr^2$ from Eqs. (\ref{Eq_r}) and (\ref{Eq_drdrho}) in Eq. (\ref{Eq_linear})(which has a far-field or slow rotation approximation) we obtain as

\begin{widetext}
    \begin{equation}
\begin{split}  
    ds^2=&\left[1-\frac{r_g}{\rho(1+\frac{r_g-2b}{4\rho})^2+2b} +\frac{2r_g\alpha}{\rho(1+\frac{r_g-2b}{4\rho})^2+2b}\frac{d\phi}{cdt}\right]c^2dt^2 
    -\frac{(1-\frac{(r_g-2b)^2}{16\rho^2})^2}{1-\frac{r_g}{\rho(1+\frac{r_g-2b}{4\rho})^2+2b}}d\rho^2 -\rho(1+\frac{r_g-2b}{4\rho})^2\\
    &\times \left\{ \rho(1+\frac{r_g-2b}{4\rho})^2+2b\right\}\left(d\theta^2+d\phi^2\right). \\
\end{split}    
\end{equation}
\end{widetext}
Thus, by simplifying the isotropic form of the KSBH solution becomes
\begin{widetext}
    \begin{equation}\label{Eq_isotropic}
\begin{split}
    ds^2= &\left[\frac{(1-\frac{r_g-2b}{4\rho})^2+2r_g\frac{\alpha}{\rho
    }\frac{d\phi}{cdt}}{(1+\frac{r_g-2b}{4\rho})^2+\frac{2b}{\rho}}\right]c^2dt^2 
    -\left\{1+\frac{r_g-2b}{4\rho}\right\}^2\left[(1+\frac{r_g-2b}{4\rho})^2+\frac{2b}{\rho}\right]d\rho^2 \\
    & -\rho^2\left\{1+\frac{r_g-2b}{4\rho}\right\}^2\left[(1+\frac{r_g-2b}{4\rho})^2+\frac{2b}{\rho}\right](d\theta^2+d\phi^2) \\
     = &\left[\frac{(1-\frac{r_g-2b}{4\rho})^2+2r_g\frac{\alpha}{\rho}\frac{d\phi}{cdt}}{(1+\frac{r_g-2b}{4\rho})^2+\frac{2b}{\rho}}\right]c^2dt^2 
     - \left\{1+\frac{r_g-2b}{4\rho}\right\}^2\left[(1+\frac{r_g-2b}{4\rho})^2+\frac{2b}{\rho}\right] \times \left[d\rho^2+\rho^2(d\theta^2+d\phi^2)\right].
\end{split}    
\end{equation}
\end{widetext}
From the isotropic form of the solution, it is possible to obtain the refractive index of the effective medium by using the material medium approach.
By setting $ds = 0$, the velocity of light ($v(\rho,\alpha,b)$) can be identified from the expression of Eq.(\ref{Eq_isotropic}) as:
\begin{equation} \label{Eq_vel-rho}
v(\rho,\alpha,b)=\frac{\sqrt{(1-\frac{r_g-2b}{4\rho})^2+2r_g\frac{\alpha}{\rho}\frac{d\phi}{cdt}}}{(1+\frac{r_g-2b}{4\rho})\left[(1+\frac{r_g-2b}{4\rho})^2+\frac{2b}{\rho}\right]}c.
\end{equation}
But this expression of the velocity of light is in the unit of length $\rho$ per unit of time. So, in order to express the velocity in terms of $r$ per unit time, we write (using Eq.(\ref{Eq_drdrho}) and (\ref{Eq_vel-rho}))
\begin{equation} 
\begin{split}    v(r,\alpha,b)&=v(\rho,\alpha,b)\frac{dr}{d\rho} \\
    &=\frac{\sqrt{(1-\frac{r_g-2b}{4\rho})^2+2r_g\frac{\alpha}{\rho}\frac{d\phi}{cdt}}}{(1+\frac{r_g-2b}{4\rho})\left[(1+\frac{r_g-2b}{4\rho})^2+\frac{2b}{\rho}\right]}\\
    &\times \left[1-(\frac{r_g-2b}{4\rho})^2\right] c\\
    &=\frac{[4\rho-(r_g-2b)]^2}{[4\rho+(r_g-2b)]^2+32b\rho}\\
    &\times \sqrt{1+8 r_g \alpha\frac{d\phi}{cdt}\frac{4\rho }{[4\rho-(r_g-2b)]^2}} c.
\end{split}
\end{equation}
Substituting the value of $\rho$ from Eq.(\ref{Eq_rho}), velocity of light becomes
\begin{equation}\label{Eq_vel_r}
    v(r,\alpha,b)= \frac{r-(r_g-2b)}{r+2b}\sqrt{1+\frac{2 \alpha r_g}{r-(r_g-2b)}\frac{d\phi}{c dt}} c.\\    
\end{equation}
Thus the refractive index of the effective medium is
\begin{equation}\label{Eq_ref_r}
    n(r,\alpha,b)= \frac{r+2b}{r-(r_g-2b)}\left[1+\frac{2 \alpha r_g}{r-(r_g-2b)} \frac{d\phi}{c dt}\right] ^{-\frac{1}{2}}.\\    
\end{equation}
Now, by replacing $\frac{r}{r_g}$ by $x$ and $\frac{b}{r_g}$ by $l$, we can write the above equation of the refractive index $ n(x,\theta,l)$ as follows
\begin{equation} \label{Eq_ref_x}
    n(x,\alpha,l)=\frac{x+2l}{x-(1-2l)}\left[1+\frac{2 \alpha}{x-(1-2l)}\frac{d\phi}{c dt}\right]^{-\frac{1}{2}},  
\end{equation}
where, if $l=0$ (i.e. $b=0$) we find that gravitational mass becomes electrically neutral and therefore the above expression of refractive index (\ref{Eq_ref_x})  goes to that for KBH metric \cite{roy2015trajectory} . Further more, if $\alpha$ is set to zero then the above expression shifts to that of the SBH case \cite{sen2010more}. 
The frame-dragging parameter, $\frac{d\phi}{cdt}$ in the second term of Eq. (\ref{Eq_ref_x}),  arises as a result of the rotation of the gravitating mass. For a slow rotating body, Eq. (\ref{Eq_ref_x}) can be expanded into an infinite converging series as it is significantly less than 1.

\subsection{Frame-dragging}
The relativistic action function $S$, for a particle with time $t$ and angle $\phi$ as cyclic variables, in the gravitational field of a rotating spherical mass\cite{landau2013classical}, is as follows
\begin{equation} \label{Eq_action}
    S=-E_0t+L\phi+S_r(r)+S_\theta(\theta)
\end{equation}
In the above equation $E_0$ denotes the conserved energy, while $L$ represents the part of the angular momentum along the field's symmetry axis and $S_r$, $S_\theta$ represent the parts of the action associated with the radius and the polar angle respectively.\\ 
Now, the four momentum of the particle can be written as:
\begin{equation} \label{Eq_four_momentum}
     p^i=mc\frac{dx^i}{ds}=g^{ik}p_k=-g^{ik}\frac{\partial S}{\partial x^k}.
\end{equation}
Here, i and k have values that go from 0 to 3 which stand for
the coordinates $ct$, $r$, $\theta$ , $\phi$ respectively \cite{landau2013classical}. 
To obtain the value of frame dragging in the case of KSBH, we have generated the following equations from Eq. (\ref{Eq_four_momentum})
\begin{equation}
\begin{split}
    mc^2\frac{dt}{ds}=&-\frac{1}{\Delta}\left[r(r+2b)+\alpha^2+\frac{r_gr\alpha^2sin^2{\theta}}{\Sigma^2}\right]\left(-\frac{E_0}{c}\right)\\
    &-\left(\frac{r_gr\alpha}{\Sigma^2 \Delta}\right)L,
\end{split}
\end{equation}
\\
and,
\begin{equation}
    mc\frac{d\phi}{ds}=-\left(\frac{r_gr\alpha}{\Sigma^2 \Delta}\right)\left(-\frac{E_0}{c}\right)+\frac{\Sigma^2-r_gr}{\Sigma^2 \Delta sin^2{\theta}}L.
\end{equation}
\\
So, the value of frame dragging ($\frac{d\phi}{cdt}$)) is obtained as
\begin{widetext}
    \begin{equation} \label{Eq:dphi_cdt}
    \frac{d\phi}{cdt}=\frac{r_gr\alpha sin^2{\theta} \frac{E_0}{c}+(\Sigma^2-r_gr)L}{\left[\left\{\Sigma^2(r(r+2b)+\alpha^2)+r_gr\alpha^2 sin^2{\theta}\right\}\frac{E_0}{c}-r_gr\alpha L\right]sin^2{\theta}}.
\end{equation}
\end{widetext}
This is the general expression of the frame-dragging for KSBH.
\begin{figure*}
	\begin{center}
    \begin{subfigure}[]
     {\includegraphics[width=0.45\textwidth]{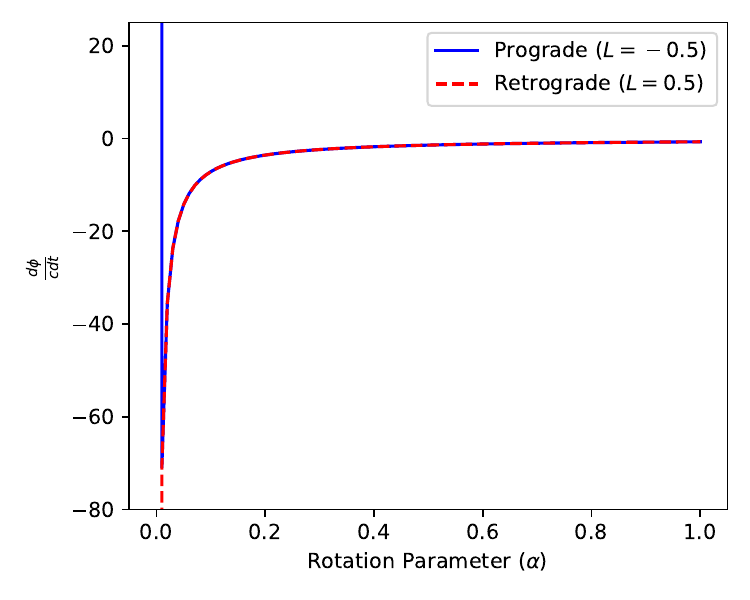}}   
    \end{subfigure}
     \begin{subfigure}[]
     {\includegraphics[width=0.45\textwidth]{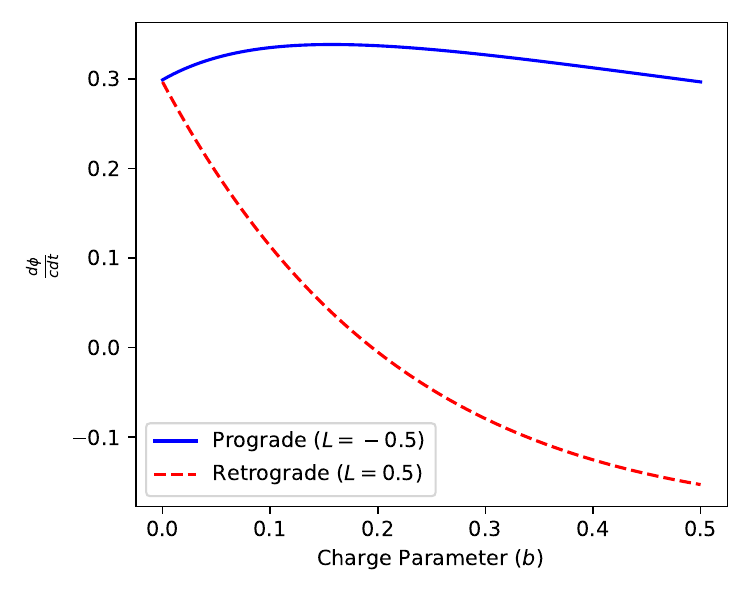}} 
    \end{subfigure}   
    \begin{subfigure}[]
     {\includegraphics[width=0.45\textwidth]{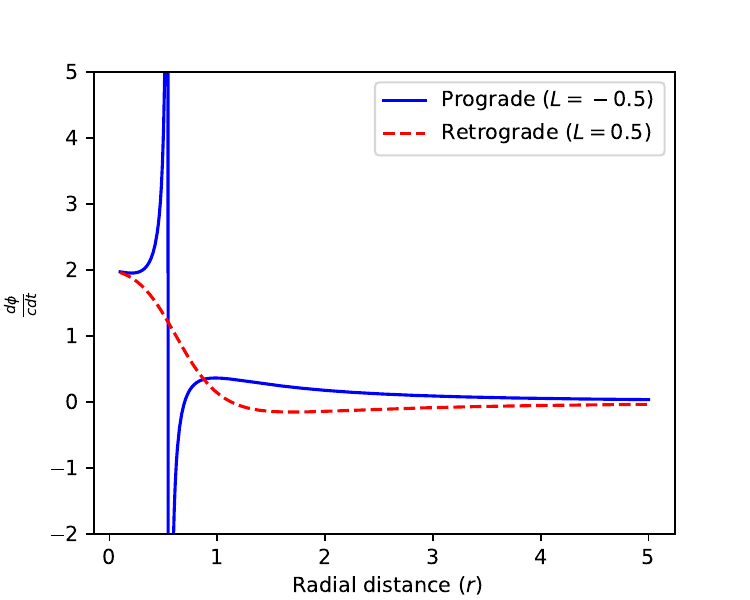}} 
    \end{subfigure} 
    \begin{subfigure}[]
     {\includegraphics[width=0.45\textwidth]{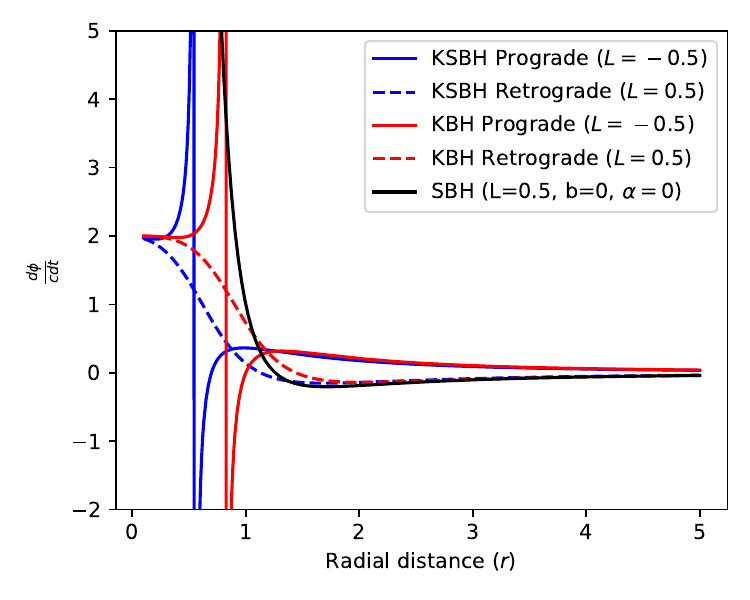}} 
    \end{subfigure} 
    \begin{subfigure}[]
     {\includegraphics[width=0.45\textwidth]{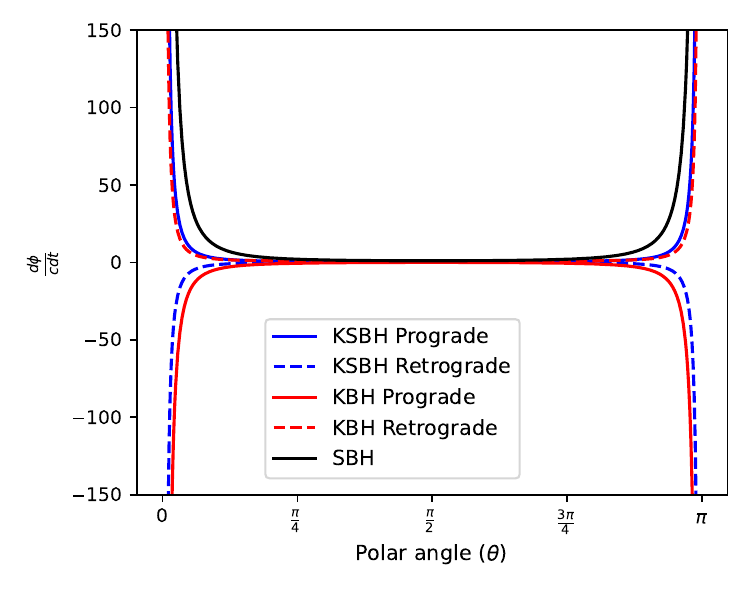}} 
    \end{subfigure}
    \begin{subfigure}[]
     {\includegraphics[width=0.45\textwidth]{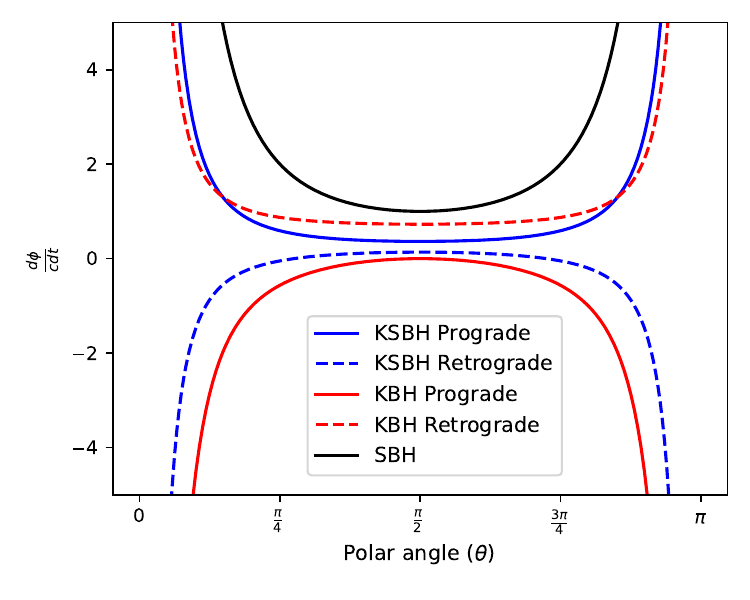}} 
    \end{subfigure}
	\end{center}
	\caption{The variation of frame dragging with various parameter and its comparison with other BHs in GR. Sub-figure (f) is the zoom in view of sub-figure (e). Here we consider $M=1$ and $E_0=-0.5$. We used $b=0.25$, $\alpha=0.5$, $\theta=\pi/2$ and $r$ near the horizon, where frame dragging not plotted as a function of these parameters.} \label{FDfig}
\end{figure*}

We started by considering the broad three-dimensional form of the KSBH. However, later we restricted our geometry to the equatorial plane with the choice of $\theta=\frac{\pi}{2}$ with the far-field approximation. Now we shall apply the same restriction to the generic expression for frame-dragging, expressed by Eq. (\ref{Eq:dphi_cdt}).
By restricting the light ray to the equatorial plane $(\theta=\frac{\pi}{2})$, the angular momentum $L$ can be expressed as
$L=p\beta  $,
where $\beta$ is the impact parameter
and for light-like particles (photons), the momentum ($p$) and the conserved energy ($E_0$) are related as $E_0=pc $
which yields $\beta=\frac{Lc}{E_0}$.  

Now, for the equatorial plane, $\Sigma^2=r(r+2b)$. Thus, Eq. (\ref{Eq:dphi_cdt}) may be written as
\begin{equation} 
      \frac{d\phi}{cdt}=\frac{r_gr\alpha+(\Sigma^2-r_gr)\beta}{\Sigma^4(1+\frac{\alpha^2}{\Sigma^2}+\frac{r_gr\alpha^2}{\Sigma^4})-r_gr\alpha\beta}
\end{equation}
Now, by far field approximation the above equation becomes
\begin{equation} \label{Eq_dphidt}
\begin{split}
    \frac{d\phi}{c dt}&=\frac{r_g r\alpha+\{r(r+2b)-r_g\}\beta}{r^2(r+2b)^2-r_g\alpha\beta}\\
    &=\frac{r_g\alpha+\{r(r+2b)-r_g\}\beta}{r(r+2b)^2-r_g\alpha\beta}\\
    &=\frac{u+(x+2l-1)v}{r_g[x(x+2l)^2-u v]}
\end{split}
\end{equation}
where $\frac{r}{r_g}=x$, $\frac{b}{r_g}=l$ (as before), $\frac{\alpha}{r_g}=u$ and $\frac{\beta}{r_g}=v$.\\
By substituting the value of $\frac{d\phi}{cdt}$ from Eq.(\ref{Eq_dphidt}) into Eq. (\ref{Eq_vel_r}), the velocity of propagation of a light ray in KSBH geometry can be expressed as:
\begin{widetext}
    \begin{equation} 
v(r,\alpha,b) = \frac{r-(r_g-2b)}{r+2b}\left[1+\frac{2 \alpha r_g}{r-(r_g-2b)}\frac{r_g\alpha+\{r(r+2b)-r_g\}\beta}{r(r+2b)^2-r_g\alpha\beta}\right] ^{\frac{1}{2}} c
\end{equation}
which in turn implies,
\begin{equation}  
v(x, u, l) =\left(1-\frac{1}{x}\right) \left\{1+\frac{2l}{(x+2l)(x-1)}\right\}\left[1+2u\frac{u +(x+2l-1)v}{(x+2l-1)(x(x+2l)^2-u v)}\right] ^{\frac{1}{2}} c
\end{equation}
Thus, the refractive index $n(x,u,l)$ (\ref{Eq_ref_x}) at an arbitrary point on equatorial plane in the KSBH field is:
\begin{equation}   
     n(x,u,l)=(\frac{x}{x-1})\left[1+\frac{2l}{(x+2l)(x-1)}\right]^{-1}\left[1+2u\frac{u +(x+2l-1)v}{(x+2l-1)(x(x+2l)^2-uv)}\right]^{-\frac{1}{2}} 
\end{equation}
\end{widetext}
which can be written in following simplified form as
\begin{equation} \label{Eq_ref_index}
n(x,u,l)=n_0(x)(1+C_x)^{-1}(1+2S_x)^{-\frac{1}{2}} =n_0(x).n_l.n_x 
\end{equation}
with the introduction of parameters as:
\begin{equation*}
\begin{split}
n_0(x)&=(\frac{x}{x-1}), \\
C_x&=\frac{2l}{(x+2l)(x-1)}, \\
S_x&=u\frac{u +(x+2l-1)v}{(x+2l-1)(x(x+2l)^2-uv)}, \\
n_l&=[1+C_x]^{-1}, \\
n_x&=[1+2S_x]^{-\frac{1}{2}}
\end{split}   
\end{equation*}
The term enclosed in parenthesis in the above formulation of refractive index (\ref{Eq_ref_index}) is related to SBH geometry \cite{sen2010more}. The second term represents the contribution of electric charge and the last square bracket represents the contribution due to combination of rotation and electric charge of the gravitating object.\\
With $l=0$ ($b=0$ or $Q=0$) i.e. when the gravitational mass becomes electrically neutral, the above expression of refractive index (\ref{Eq_ref_index})  goes to that for KBH geometry \cite{roy2015trajectory}.

Here we can show that for all $r>r_g$, we get $S_x<<$1 and $C_x<<$1.
This is possible, as when $r>>r_g$ and $r>>\alpha$, we must have $x>>1$ and $x>>u$. Also, since $\alpha<\beta$, we must have $u<v$.

Now, as $x >> 1$, we can approximate $(x-1) \approx
x$ and then we can finally show that $S_x << 1$ and $C_x<<$1.

In FIG.\ref{FDfig} the frame dragging effect of KSBH is illustrated as a function of rotation parameter, charge parameter, radial coordinates, and polar coordinate, respectively. In FIG. \ref{FDfig}(a) to (c) different color encode prograde and retrograde motion of KSBH. In FIG. \ref{FDfig}(d) to (e) frame dragging of KSBH is illustrated and compared with other well-known BH solutions (KBH and SBH), where different colors encode different BHs and retrograde motions are plotted with a dashed line. In the caption of this figure, it is already mentioned that Figs. (a) to (d) are plotted in the equatorial plane.  

In FIG.\ref{FDfig}(a), the frame dragging effect is more prominent in prograde motion than that of retrograde motion. SBH has no dependence on the rotation parameter and in the presence of spin on BH in the same direction of light ray, the drag is more than in the opposite direction of the light ray.
FIG.\ref{FDfig}(b) is plotted at fixed rotation parameter  $\alpha =0.5$,  
showing that frame dragging decreases with charge in retrograde and increases in prograde. However, at $b=0$ and the non-zero rotation parameter, the divergent nature of the trajectories reveals the prograde and retrograde nature of the BH, which increases significantly at higher charge values.  
FIG.\ref{FDfig}(d) is the reflector of FIG.\ref{FDfig}(c) to show all other well-known BH solutions in GR.
Like FIG. \ref{FDfig}(c), in FIG.\ref{FDfig}(d) also, the frame dragging plotted against the radial distances with $b=0.25$ and $\alpha=0.5$.  In the absence of the charge parameter (b=0), the frame dragging coincides with that of KBH and in the absence of the charge parameter and rotation parameter (b=0 and $\alpha=0$), the frame dragging corresponds to that of SBH.
Prograde trajectories usually exhibit a higher drag than retrograde ones, and in the presence of charge, dragging is more prominent.
Figs.\ref{FDfig}(e) and (f) reveal that the dragging of BH is more significant near the pole and the anti-pole position for all BH solutions in GR. The contrasting behavior of frame-dragging with respect to the polar angle in the KSBH compared to that of the KBH is attributed to the presence of dilaton and axion fields, which modify the underlying spacetime geometry. All BH models are more distinctive in smaller radial distances and polar angles, indicating their independence in the BH parameters.
\begin{figure}
	\begin{center}
        {\includegraphics[width=0.4\textwidth]{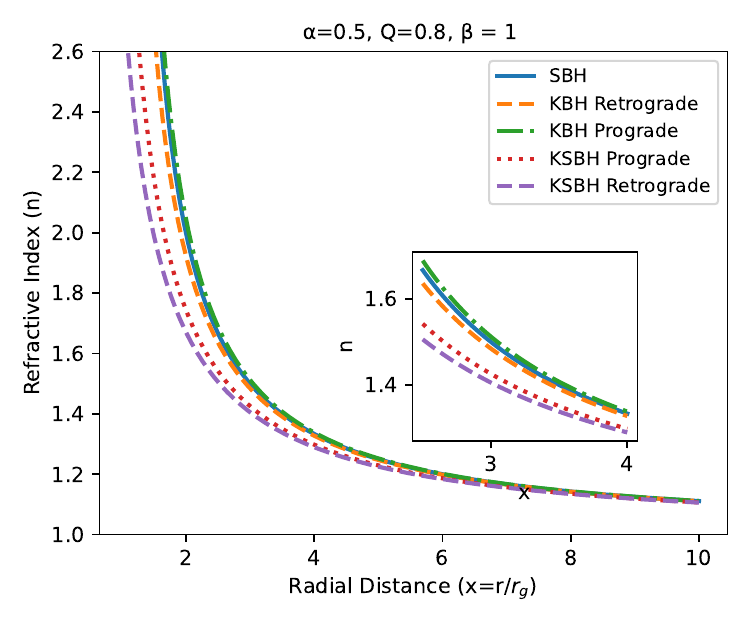}} 
	\end{center}
	\caption{The variation of refractive index with radial distance.} \label{nvsx}
\end{figure}

\begin{figure}
	\begin{center}  
        {\includegraphics[width=0.4\textwidth]{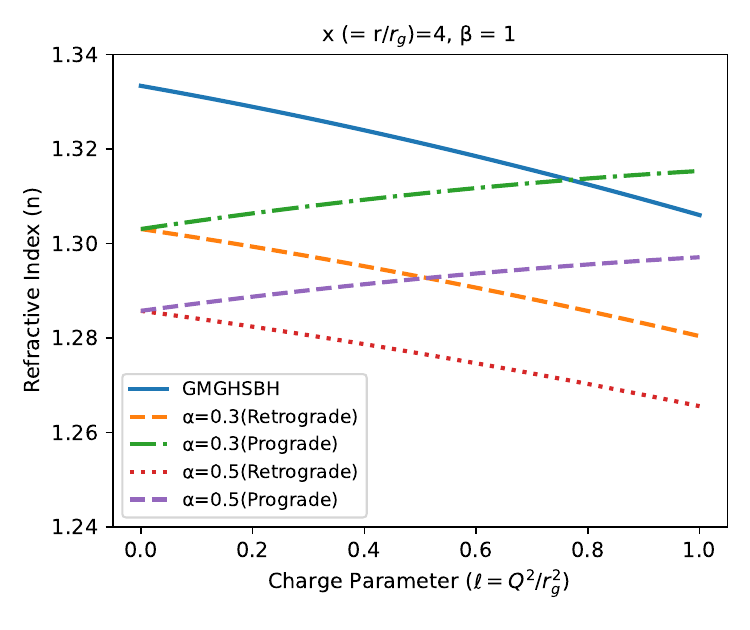}}
	\end{center}
	\caption{The variation of refractive index with charge parameter.}  \label{nvsQ}
\end{figure}

\begin{figure}
	\begin{center}       
        {\includegraphics[width=0.4\textwidth]{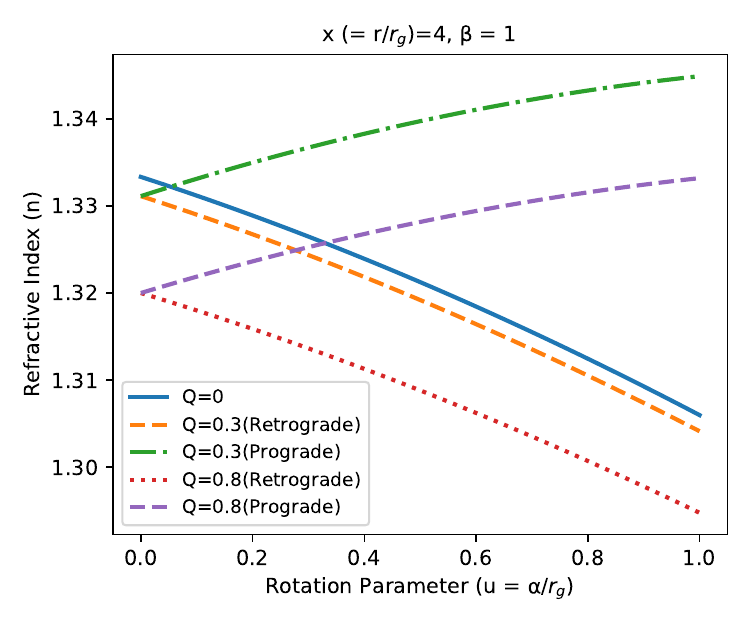}}
	\end{center}
	\caption{The variation of refractive index with rotation parameter.} \label{nvsα}
\end{figure}
\figurename{\ref{nvsx}} to \figurename{\ref{nvsα}} illustrate the variation of the refractive index (\(n\)) as a function of radial distance, charge parameter, and spin parameter respectively. In \figurename{\ref{nvsx}}, the refractive index decreases with increasing radial distance (\(x = r/r_g\)) for different spacetime geometries, with prograde trajectories consistently having higher values of \(n\) than retrograde ones. The distinctions between spacetime models (e.g., SBH, KBH and KSBH) are more pronounced at smaller radial distances. \figurename{\ref{nvsQ}} examines the dependence of \(n\) on the charge parameter (\(\ell = Q^2/r_g^2\)) at a fixed \(x = 4\), showing that \(n\) generally decreases with \(\ell\), but retrograde and prograde trajectories diverge significantly when the rotation parameter (\(\alpha\)) is non-zero. In the absence of the rotation parameter, the BH solution reduces to the GMGHSBH and the refractive index exhibits a decreasing trend with increasing charge. Further, \figurename{\ref{nvsα}} highlights the influence of the rotation parameter (\(u = \alpha/r_g\)), demonstrating that increasing spin amplifies the differences in refractive index between prograde and retrograde trajectories, especially for higher charge values. In particular we can say that the refractive index decreases with radial distance regardless of spacetime geometry. Prograde trajectories consistently exhibit higher refractive indices than retrograde ones across all cases. Differences between spacetime geometries and trajectory types become more pronounced for larger charge and spin parameters, highlighting their combined influence on the refractive index.
\begin{figure}[H] 
	\begin{center}
    \begin{subfigure}[]
     {\includegraphics[width=0.4\textwidth]{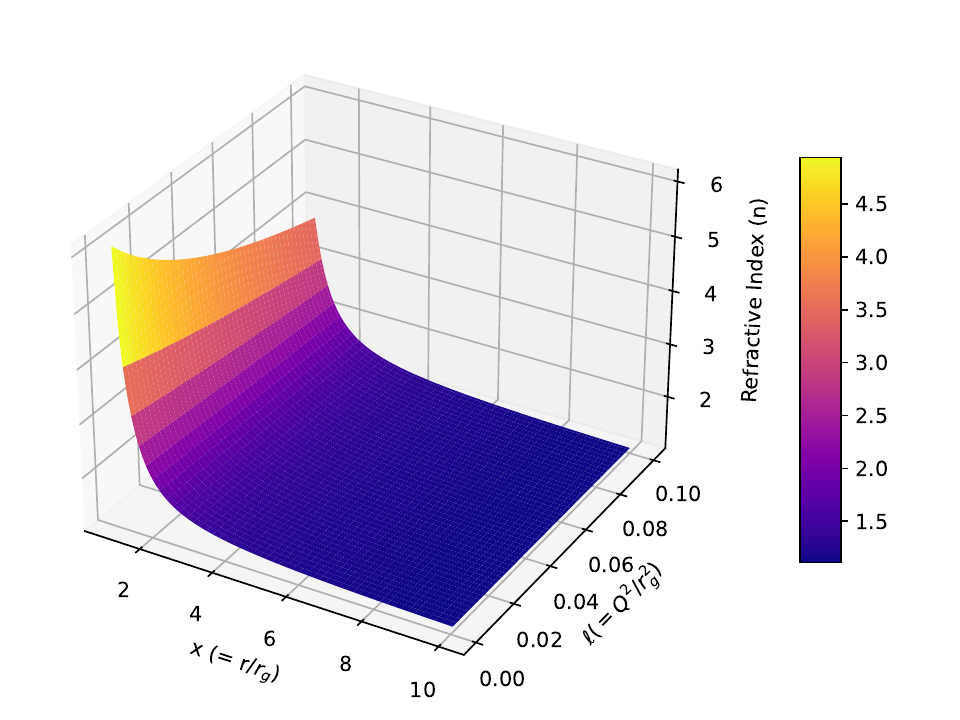}}   
    \end{subfigure}
     \begin{subfigure}[]
     {\includegraphics[width=0.4\textwidth]{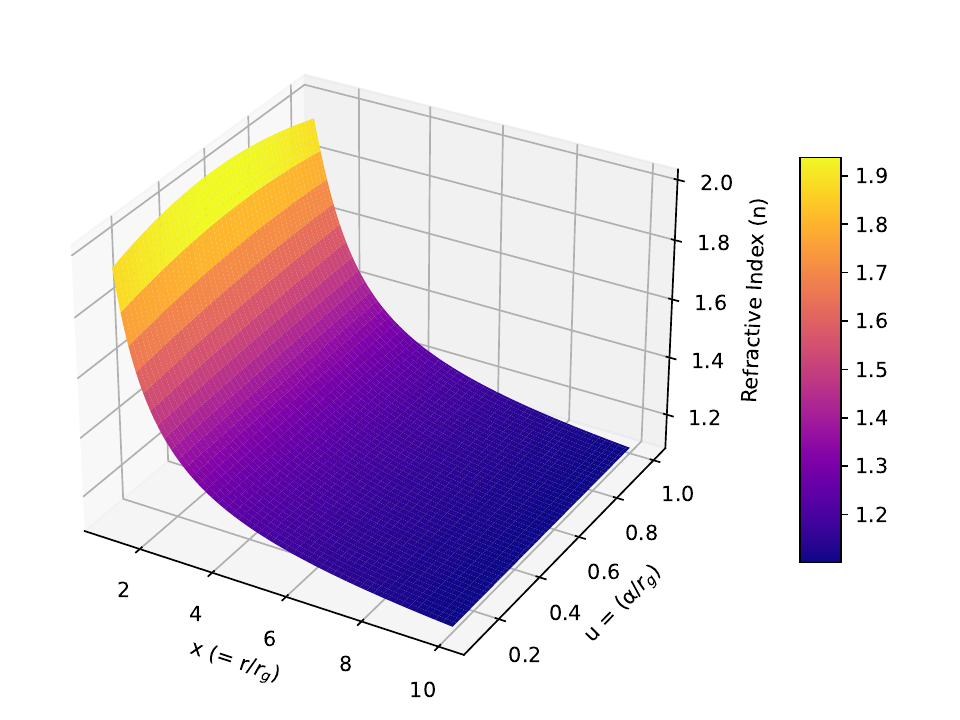}}  
    \end{subfigure}   
	\end{center}
	\caption{The three dimensional view of variation of refractive index with (a) radial distance and charge parameter with fixed rotation parameter i.e., $u =0.5$ and (b) radial distance and rotation parameter with fixed charge parameter i.e., $\ell =0.5$ }. \label{3D1}
\end{figure}
\vspace{-0.5cm}
\noindent The \figurename{\ref{3D1}} contains two three dimensional surface plots that depict the variation of the refractive index as a function with radial distance and other parameters. The $3D$ representation provides insights into how the parameters interact and affect refractive index simultaneously, offering a richer and more complete understanding of the system under study. In both plots, the refractive index decreases as the radial distance increases. The charge and rotation parameters affect the refractive index more significantly at smaller distances, while their influence diminishes at larger distances. Further, higher values of the rotation parameter result in higher values of the refractive index, especially at smaller radial distances.

\section{Deflection due to Kerr-Sen BH}
The curved space-time is related to a special optical medium with graded refractive index. Since only vacuum exists between the gravitational matter, it may be considered that vacuum is a special optical medium whose refractive index is influenced by the gravitational field. Thus, the gravitational effect is
represented by an equivalent refractive index of the medium. Therefore, if a ray of light passes through a material medium, the light ray will deviate due to the variation of the refractive index of the corresponding media. \\
Now, the deflection of the light ray and hence the trajectory of light ray depends on the refractive index of the medium, can be represented as \cite{born2013principles, sen2010more, roy2015trajectory, roy2017deflection} 
\begin{equation} \label{Eq_def}
    \Delta \psi=2\int_{\beta}^{\infty}\frac{dr}{r\sqrt{(\frac{n(r).r}{n(\beta).\beta})^2-1}}.
\end{equation}
In the present case, light approaches from asymptotic infinity ($r =- \infty$ or $x =- \infty$) towards a rotating gravitational mass at the origin, defined by the SBH radius $r_g$, rotation parameter $\alpha$ and associated charge $b$.
The beam reaches $r = \infty$ or $x = \infty$ after a specified amount of deflection ($\Delta\psi$). The oncoming ray's closest approach distance is $\beta$.(In practice, the impact parameter and the approach distance are roughly equal.) The trajectory's tangent becomes perpendicular to the vector $r$, which is $\beta$, when the light beam passes through the closest distance of approach, or $r = \beta$.
Now, we change the variable to $x = \frac{r}{r_g}$ so that $dr = r_gdx$ so that the corresponding limit changes from $x = v$ to $x =\infty$, as the limit of $r$ changes from $r = \beta$ to $r =\infty$ in the Eq. (\ref{Eq_def})
\begin{equation} \label{Eq_def_2}
\begin{split}
     \Delta \psi&=2\int_{v}^{\infty}\frac{dx}{x\sqrt{(\frac{n(x).x}{n(v).v})^2-1}}-\pi \\
    &=2I-\pi,
\end{split}
\end{equation}
where 
\begin{equation}
\begin{split}    
    I&=\int_{v}^{\infty}\frac{dx}{x\sqrt{(\frac{n(x).x}{n(v).v})^2-1}}\\
    &= n(v).v\int_{v}^{\infty}\frac{dx}{x\sqrt{(n(x).x)^2-(n(v).v)^2}}.
\end{split}
\end{equation}
Substituting the value of refractive index from Eq.(\ref{Eq_ref_index}) and $n(v).v=D_{KS}$, we rewrite the above Eq. of $I$ as 
\begin{widetext}
\begin{equation} 
\begin{split}
    I &=  D_{KS}\int_{v}^{\infty}\frac{dx}{x\sqrt{[n_0(x).x(1+C_x)^{-1}(1+2S_x)^{-\frac{1}{2}}]^2-D^2_{KS}}} \\
    &=D_{KS}\int_{v}^{\infty}\frac{dx}{x\sqrt{n^2_0(x).x^2(1+C_x)^{-2}(1+2S_x)^{-1}-D^2_{KS}}} \\
    &=D_{KS}\int_{v}^{\infty}\frac{dx}{x\sqrt{n^2_0(x).x^2-D^2_0+n^2_0(x).x^2(1+C_x)^{-2}(1+2S_x)^{-1}-n^2_0(x).x^2+D^2_0-D^2_{KS}}} \\
    &=D_{KS}\int_{v}^{\infty}\frac{dx}{x\sqrt{n^2_0(x).x^2-D^2_0}}[1+\frac{n^2_0(x).x^2[(1+C_x)^{-2}(1+2S_x)^{-1}-1]+D^2_0-D^2_{KS}}{n^2_0(x).x^2-D^2_0}]^{-\frac{1}{2}} \\
    &=D_{KS}\int_{v}^{\infty}\frac{dx}{x\sqrt{n^2_0(x).x^2-D^2_0}}[1+A(x)]^{-\frac{1}{2}}.
\end{split}
\end{equation} 
\end{widetext}
Here, $D_{0}=n_{0}(v).v$, corresponding to SBH deflection and also we have denoted
\begin{equation}
    A(x)= \frac{n^2_0(x).x^2[(1+C_x)^{-2}(1+2S_x)^{-1}-1]+D^2_0-D^2_{KS}}{n^2_0(x).x^2-D^2_0}.
\end{equation}

\begin{figure}[H]
	\begin{center}
    \begin{subfigure}[]
     {\includegraphics[width=0.45\textwidth]{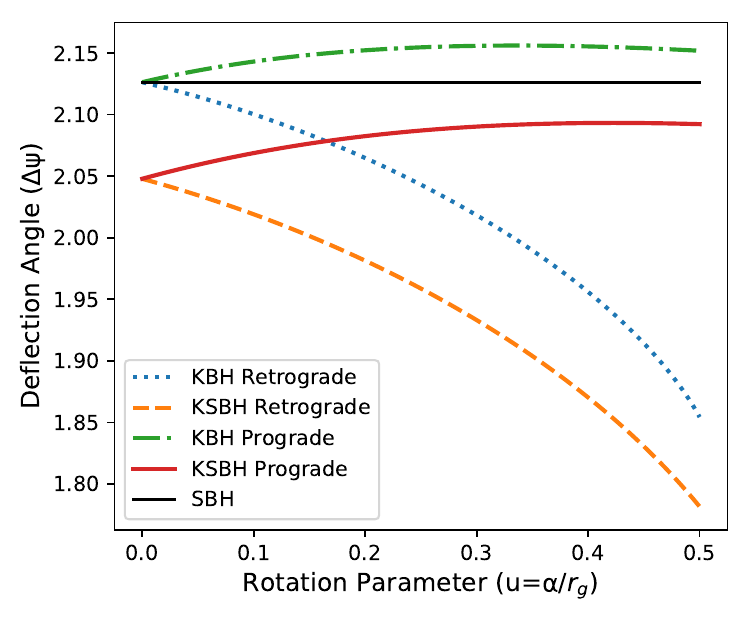}}   
    \end{subfigure}
     \begin{subfigure}[]
     {\includegraphics[width=0.45\textwidth]{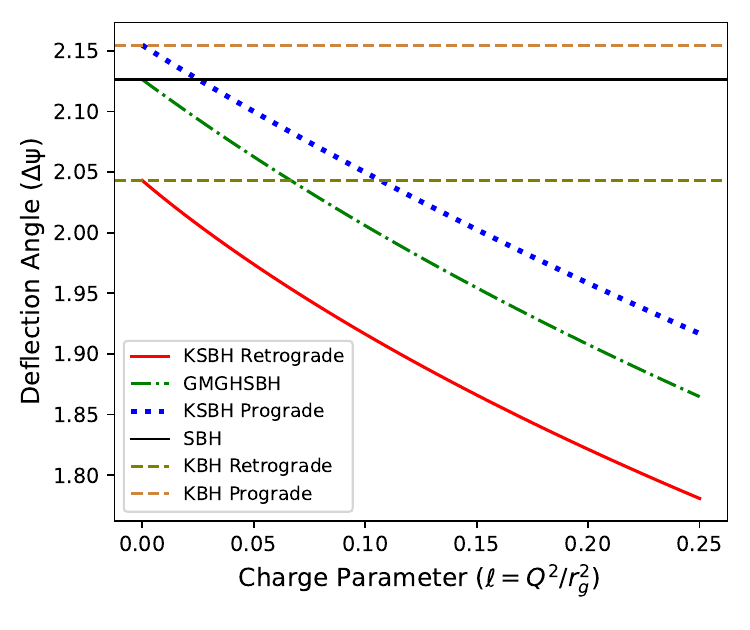}} 
    \end{subfigure}   
	\end{center}
	\caption{The variation of deflection angle with retro-grade and pro-grade motion with varying (a) rotation parameter with $\ell=0.0625$  and (b) charge parameter with $u=0.25$.} \label{DAN}
\end{figure}

\begin{figure}[H]
	\begin{center}       
        {\includegraphics[width=0.5\textwidth]{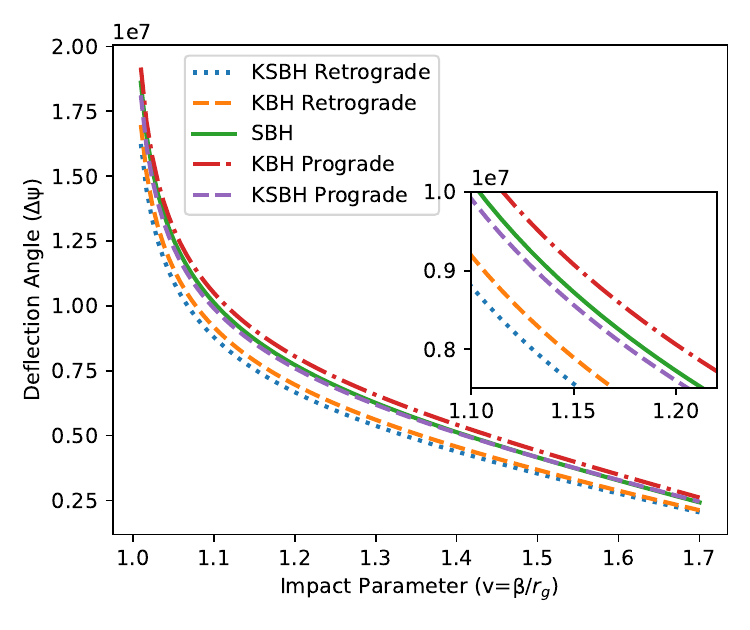}}
	\end{center}
	\caption{The variation of deflection angle with impact parameter. The plots also show the results for the Schwarzschild, Kerr, and Kerr-Sen BHs for comparison.} \label{daip}
\end{figure}

Eq. (\ref{Eq_def_2}) may be used to find the exact deflection of light in the equatorial plane of a KSBH as
\begin{equation} \label{Eq_def_final}
\begin{split}
     \Delta \psi&=2I-\pi\\
     &=2D_{KS}\int_{v}^{\infty}\frac{dx}{x\sqrt{n^2_0(x).x^2-D^2_0}}[1+A(x)]^{-\frac{1}{2}} -\pi.
\end{split}
\end{equation}
The above deflection angle can be expanded as a series
\begin{widetext}
\begin{equation} \label{Eq_def_final1}
     \Delta \psi =2D_{KS}\int_{v}^{\infty}\frac{dx}{x\sqrt{n^2_0(x).x^2-D^2_0}}[1-\frac{1}{2}A(x)+\frac{3}{8}A^2(x)-\frac{5}{16}A^3(x)+--------] -\pi.
\end{equation}
\end{widetext}
The variation of the deflection angle with the spin and charge parameters is illustrated in \figurename{\ref{DAN}}. The graphical representation also provides a comparison of the KSBH with other well-known BHs solution in GR, namely the KBH and the SBH. Our analysis shows that the deflection angle decreases with an increasing spin parameter for retrograde motion, while it increases for prograde motion. For retrograde motion, photons move against the spin-induced frame-dragging effect, experiencing reduced curvature and thus a smaller deflection angle. Conversely, in prograde motion, photons are dragged along with the spin, resulting in greater curvature and an increased deflection angle. The straight line represents the spin-independent nature of the SBH. In addition, the variation of the deflection angle with the charge parameter was analyzed in the absence of rotation, corresponding to the GMGHSBH. It was observed that the deflection angle decreases with increasing charge. Further, we observed that the deflection angle decreases with an increasing charge parameter for both trajectories. Notably, the presence of the charge parameter alters the behavior of the deflection angle for prograde motion, making it distinct from that of the KBH. However, for a specific charge value, the deflection angle is maximized for prograde orbits and reduced for retrograde orbits. This behavior arises from the charge parameter modifying the gravitational and electromagnetic interactions around the BH, amplifying the frame-dragging effect for prograde orbits and diminishing it for retrograde orbits. This highlights the intricate interplay between charge, spin, and photon dynamics in the vicinity of KSBH.\\
Furthermore, the graphical representation of deflection angle with impact parameter alongwith the comparison of well-known BHs solution can be clearly seen in \figurename{\ref{daip}}. The deflection angle for all BH solutions decreases with an increasing impact parameter, indicating its independence BH parameters. This behavior remains consistent across various BH types viz., charged, rotating, non-rotating, or charged rotating also for both prograde and retrograde motions. We also observed that, among all the studied BHs solution, the deflection angle is most pronounced for Kerr prograde motion and least pronounced for KS retrograde motion. The difference in deflection angles between Kerr and KS retrograde motions highlights the influence of charge on spacetime geometry, with the charge in the KSBH reducing the deflection angle compared to the KBH. This behavior provides valuable insight into the effects of spin and charge on photon trajectories, aiding in the distinction of BH models through observational data. One can verify our results by comparing the deflection angles obtained through null geodesics, which clearly demonstrate that the results presented in this work align well with those reported by other authors; for further details, please refer to the references cited herein \cite{Chandrasekhar1998,Claudel:2000yi,Iyer:2009wa,uniyal2018bending,Hsiao:2019ohy}.
\begin{figure}[H]
	\begin{center}       
        {\includegraphics[width=0.5\textwidth, height=8cm]{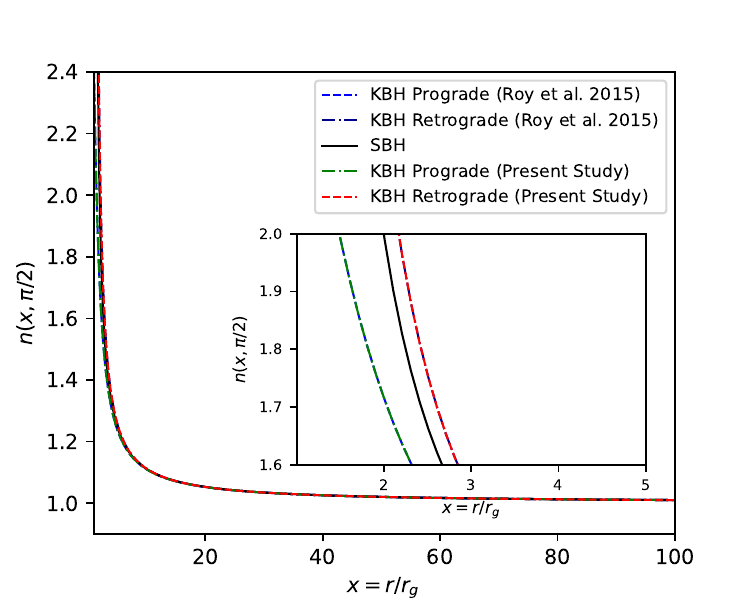}}
	\end{center}
	\caption{This plot compares the results of Roy et al. 2015 with the present case when the charge is set to zero. The results for both retrograde and prograde motion match exactly with those of Roy et al. 2015. } \label{royetal2015}
\end{figure} 
In respective limit $\alpha=0$, the KSBH reduces to the KBH, and our results exactly match those of Roy et. al.\cite{roy2015trajectory}, as illustrated in FIG. \ref{royetal2015}. Roy et. al. demonstrated that the material medium approach yields deflection angles consistent with those obtained through the conventional null geodesic method. Furthermore, their findings were verified using the Sun as a test case and by considering some millisecond pulsars to more objectively analyze the effect of rotation on the deflection angle. Therefore, we do not repeat these analyses in the present study. However, our study extends to charged rotating BHs, offering a framework to analyze deflection angles under the influence of both spin and charge parameters. These findings can be constrained with observational results from the Event Horizon Telescope (EHT) data, providing deeper insights into the interplay between BH parameters and photon trajectories near astrophysical BHs.
\section{Result and Discussion}
\noindent In this work, we study the deflection angle of light around a KSBH, using the material medium approach and compare its behavior with other well-known solutions in GR, such as the KBH and SBH.

\begin{itemize}
    \item The study of photon sphere shows that the charge parameter increases the radius of the prograde photon sphere while decreasing that of the retrograde one, showing its asymmetric influence on photon orbits. Conversely, the rotation parameter causes the retrograde photon sphere to expand and the prograde to shrink, reflecting their distinct roles in shaping the photon sphere structure.
    
    \item The contrasting behavior of frame-dragging with respect to the polar angle in the KSBH compared to the KBH is attributed to the presence of dilaton and axion fields, which modify the underlying spacetime geometry. This highlights the impact of string-theoretic corrections on rotational effects in gravitational regime. Such deviations could serve as potential observational signatures that distinguish classical and string-inspired BHs.
    
    \item The deflection angle exhibits a distinct dependence on the spin parameter, decreasing for retrograde motion and increasing for prograde motion. This variation arises from the spin-induced frame-dragging effect, with photons in retrograde motion encountering reduced spacetime curvature.
    
    \item The inclusion of the charge parameter modifies the deflection behavior. For a specific charge value, the deflection angle is maximized for prograde orbits and minimized for retrograde orbits, highlighting the interplay between gravitational and electromagnetic interactions.

    \item The deflection angle decreases with increasing impact parameter across all studied BH solutions, reflecting its independence from the intrinsic parameters of BHs, such as charge or spin.

    \item Comparative analysis reveals that the KBH exhibits the most pronounced deflection angle for prograde motion, whereas the KSBH shows the least deflection angle for retrograde motion due to charge-induced suppression of spacetime curvature.

    \item In the respective limit $\alpha = 0$, the KSBH reduces to KBH, and the obtained results precisely align with the findings of Roy et al.\cite{roy2015trajectory}, which validated deflection angles through both the material medium and the null geodesic approaches.

    \item In the respective limits $\alpha = 0$ and $b=0$, the KSBH reduces to the well-known SBH solution in GR, and the results obtained precisely align with the findings of Sen\cite{sen2010more}, which again strongly validated the deflection angles through both the material medium and the null geodesic approaches.

    \item The extended framework for charged rotating BHs provides a robust methodology to explore the combined effects of spin and charge on photon trajectories. These findings hold significant potential for constraining theoretical models with observational data from the EHT Collaboration.
\end{itemize}
	
\begin{acknowledgments}
Authors, SR and AS express sincere and deep gratitude to the Department of Physics, NITA, for providing the necessary research environment to complete this work. 
The author, SK, sincerely acknowledges IMSc for providing exceptional research facilities and a conducive environment that facilitated his work as an Institute Postdoctoral Fellow. One of the authors, HN, would like to thank IUCAA, Pune, for the support under its associateship program, where a part of this work was done. The author HN also acknowledges the financial support provided by the Science and Engineering Research Board (SERB), New Delhi, through grant number CRG/2023/008980.
\end{acknowledgments}
\newpage
\appendix
\section{Photon Sphere Expressions}
The equation for the radii of the photon sphere, presented in Eq.~\ref{PSEq}, contains undefined symbols $\mathcal{A}$ and $\mathcal{B}$ due to the complexity of their mathematical expressions. To enhance clarity, we provide the explicit expressions for these symbols in this appendix. The term $\mathcal{A}$ is given by,
\begin{equation}
    \mathcal{A} = \frac{1}{2} \sqrt{\left( \frac{36 M^{2}+25 b^{2}+20Mb}{12} + \frac{1}{12} \left( \chi + \frac{\sigma} \chi \right)\right)},
\end{equation}
where
\begin{equation}
    \chi = \sqrt[3]{\frac{1}{2} \left( \Sigma +  \Sigma^{2} - 4 \sigma^{3} \right)},
\end{equation}
with
\begin{align}
    \sigma = & \left( 36 M^2 - 44 M b + 13 b^2 \right)^2 \notag \\
    & - 3b \left[ -(16 M \alpha^2 - 6b(b-2M)^2) \right] \notag \\
    & + 12\alpha \left[ -(8 M \alpha^2 - b(b-2M)^2) \right],
\end{align}
and
\begin{align}
\Sigma = & \, 2 \left( 36 M^2 - 44 M b + 13 b^2 \right)^3  \notag \\
& - 9b \left( 36 M^2 - 44 M b + 13 b^2 \right) \left[ -(16 M \alpha^2 - 6b(b-2M)^2) \right]  \notag \\
& + 27 b^2 \left[ -(8 M \alpha^2 - b(b-2M)^2) \right]  \notag \\
& + 27\alpha \left[ -(16 M \alpha^2 - 6b(b-2M)^2) \right]^2  \notag \\
& - 72\alpha \left[ -(8 M \alpha^2 - b(b-2M)^2) \right] \left( 36 M^2 - 44 M \beta + 13 b^2 \right) \notag \\
& + 27 b^2 \left[ -(8 M \alpha^2 - b(b-2M)^2) \right]  \notag \\
& + 27a \left[ -(16 M \alpha^2 - 6b(b-2M)^2) \right]^2  \notag \\
& - 72a \left[ -(8 M \alpha^2 - b(b-2M)^2) \right] \left( 36 M^2 - 44 M \beta + 13 b^2 \right).
\end{align}

Similarly, the expression for $\mathcal{B}$ is
\begin{widetext}
    \begin{equation}
    \mathcal{B} = \frac{\left[ 12 (b - 2M) \right]^3 - 16 \left[ 12 (b - 2M) \right] (36M^2 - 44Mb + 13b^2) - 2048 M \alpha^2 + 768 b (b - 2M)^2}{512}.
\end{equation}
\end{widetext}

\nocite{*}

\bibliography{KerrSen}

\begin{thebibliography}{50}%
\makeatletter
\providecommand \@ifxundefined [1]{%
 \@ifx{#1\undefined}
}%
\providecommand \@ifnum [1]{%
 \ifnum #1\expandafter \@firstoftwo
 \else \expandafter \@secondoftwo
 \fi
}%
\providecommand \@ifx [1]{%
 \ifx #1\expandafter \@firstoftwo
 \else \expandafter \@secondoftwo
 \fi
}%
\providecommand \natexlab [1]{#1}%
\providecommand \enquote  [1]{``#1''}%
\providecommand \bibnamefont  [1]{#1}%
\providecommand \bibfnamefont [1]{#1}%
\providecommand \citenamefont [1]{#1}%
\providecommand \href@noop [0]{\@secondoftwo}%
\providecommand \href [0]{\begingroup \@sanitize@url \@href}%
\providecommand \@href[1]{\@@startlink{#1}\@@href}%
\providecommand \@@href[1]{\endgroup#1\@@endlink}%
\providecommand \@sanitize@url [0]{\catcode `\\12\catcode `\$12\catcode `\&12\catcode `\#12\catcode `\^12\catcode `\_12\catcode `\%12\relax}%
\providecommand \@@startlink[1]{}%
\providecommand \@@endlink[0]{}%
\providecommand \url  [0]{\begingroup\@sanitize@url \@url }%
\providecommand \@url [1]{\endgroup\@href {#1}{\urlprefix }}%
\providecommand \urlprefix  [0]{URL }%
\providecommand \Eprint [0]{\href }%
\providecommand \doibase [0]{https://doi.org/}%
\providecommand \selectlanguage [0]{\@gobble}%
\providecommand \bibinfo  [0]{\@secondoftwo}%
\providecommand \bibfield  [0]{\@secondoftwo}%
\providecommand \translation [1]{[#1]}%
\providecommand \BibitemOpen [0]{}%
\providecommand \bibitemStop [0]{}%
\providecommand \bibitemNoStop [0]{.\EOS\space}%
\providecommand \EOS [0]{\spacefactor3000\relax}%
\providecommand \BibitemShut  [1]{\csname bibitem#1\endcsname}%
\let\auto@bib@innerbib\@empty
\bibitem [{\citenamefont {Schwarzschild}(1916)}]{schwarzschild1916gravitationsfeld}%
  \BibitemOpen
  \bibfield  {author} {\bibinfo {author} {\bibfnamefont {K.}~\bibnamefont {Schwarzschild}},\ }\bibfield  {title} {\bibinfo {title} {{\"U}ber das gravitationsfeld eines massenpunktes nach der einsteinschen theorie},\ }\href@noop {} {\bibfield  {journal} {\bibinfo  {journal} {Sitzungsberichte der k{\"o}niglich preussischen Akademie der Wissenschaften}\ ,\ \bibinfo {pages} {189}} (\bibinfo {year} {1916})}\BibitemShut {NoStop}%
\bibitem [{\citenamefont {Kerr}(1963)}]{kerr1963gravitational}%
  \BibitemOpen
  \bibfield  {author} {\bibinfo {author} {\bibfnamefont {R.~P.}\ \bibnamefont {Kerr}},\ }\bibfield  {title} {\bibinfo {title} {Gravitational field of a spinning mass as an example of algebraically special metrics},\ }\href@noop {} {\bibfield  {journal} {\bibinfo  {journal} {Physical review letters}\ }\textbf {\bibinfo {volume} {11}},\ \bibinfo {pages} {237} (\bibinfo {year} {1963})}\BibitemShut {NoStop}%
\bibitem [{\citenamefont {Reissner}(1916)}]{reissner1916eigengravitation}%
  \BibitemOpen
  \bibfield  {author} {\bibinfo {author} {\bibfnamefont {H.}~\bibnamefont {Reissner}},\ }\bibfield  {title} {\bibinfo {title} {{\"U}ber die eigengravitation des elektrischen feldes nach der einsteinschen theorie},\ }\href@noop {} {\bibfield  {journal} {\bibinfo  {journal} {Annalen der Physik}\ }\textbf {\bibinfo {volume} {355}},\ \bibinfo {pages} {106} (\bibinfo {year} {1916})}\BibitemShut {NoStop}%
\bibitem [{\citenamefont {Nordstr{\"o}m}(1918)}]{nordstrom1918energy}%
  \BibitemOpen
  \bibfield  {author} {\bibinfo {author} {\bibfnamefont {G.}~\bibnamefont {Nordstr{\"o}m}},\ }\bibfield  {title} {\bibinfo {title} {On the energy of the gravitation field in einstein’s theory},\ }\href@noop {} {\bibfield  {journal} {\bibinfo  {journal} {Koninklijke Nederlandse Akademie van Wetenschappen Proceedings Series B Physical Sciences}\ }\textbf {\bibinfo {volume} {20}},\ \bibinfo {pages} {1238} (\bibinfo {year} {1918})}\BibitemShut {NoStop}%
\bibitem [{\citenamefont {Janis}\ \emph {et~al.}(1968)\citenamefont {Janis}, \citenamefont {Newman},\ and\ \citenamefont {Winicour}}]{janis1968reality}%
  \BibitemOpen
  \bibfield  {author} {\bibinfo {author} {\bibfnamefont {A.~I.}\ \bibnamefont {Janis}}, \bibinfo {author} {\bibfnamefont {E.~T.}\ \bibnamefont {Newman}},\ and\ \bibinfo {author} {\bibfnamefont {J.}~\bibnamefont {Winicour}},\ }\bibfield  {title} {\bibinfo {title} {Reality of the schwarzschild singularity},\ }\href@noop {} {\bibfield  {journal} {\bibinfo  {journal} {Physical Review Letters}\ }\textbf {\bibinfo {volume} {20}},\ \bibinfo {pages} {878} (\bibinfo {year} {1968})}\BibitemShut {NoStop}%
\bibitem [{\citenamefont {Newman}\ \emph {et~al.}(1965)\citenamefont {Newman}, \citenamefont {Couch}, \citenamefont {Chinnapared}, \citenamefont {Exton}, \citenamefont {Prakash},\ and\ \citenamefont {Torrence}}]{newman1965metric}%
  \BibitemOpen
  \bibfield  {author} {\bibinfo {author} {\bibfnamefont {E.~T.}\ \bibnamefont {Newman}}, \bibinfo {author} {\bibfnamefont {E.}~\bibnamefont {Couch}}, \bibinfo {author} {\bibfnamefont {K.}~\bibnamefont {Chinnapared}}, \bibinfo {author} {\bibfnamefont {A.}~\bibnamefont {Exton}}, \bibinfo {author} {\bibfnamefont {A.}~\bibnamefont {Prakash}},\ and\ \bibinfo {author} {\bibfnamefont {R.}~\bibnamefont {Torrence}},\ }\bibfield  {title} {\bibinfo {title} {Metric of a rotating, charged mass},\ }\href@noop {} {\bibfield  {journal} {\bibinfo  {journal} {Journal of mathematical physics}\ }\textbf {\bibinfo {volume} {6}},\ \bibinfo {pages} {918} (\bibinfo {year} {1965})}\BibitemShut {NoStop}%
\bibitem [{\citenamefont {Fujii}\ and\ \citenamefont {Maeda}(2003)}]{fujii2003scalar}%
  \BibitemOpen
  \bibfield  {author} {\bibinfo {author} {\bibfnamefont {Y.}~\bibnamefont {Fujii}}\ and\ \bibinfo {author} {\bibfnamefont {K.-i.}\ \bibnamefont {Maeda}},\ }\href@noop {} {\emph {\bibinfo {title} {The scalar-tensor theory of gravitation}}}\ (\bibinfo  {publisher} {Cambridge University Press},\ \bibinfo {year} {2003})\BibitemShut {NoStop}%
\bibitem [{\citenamefont {Tong}(2009)}]{tong2009lectures}%
  \BibitemOpen
  \bibfield  {author} {\bibinfo {author} {\bibfnamefont {D.}~\bibnamefont {Tong}},\ }\bibfield  {title} {\bibinfo {title} {Lectures on string theory},\ }\href@noop {} {\bibfield  {journal} {\bibinfo  {journal} {arXiv preprint arXiv:0908.0333}\ } (\bibinfo {year} {2009})}\BibitemShut {NoStop}%
\bibitem [{\citenamefont {SenGupta}(2008)}]{sengupta2008aspects}%
  \BibitemOpen
  \bibfield  {author} {\bibinfo {author} {\bibfnamefont {S.}~\bibnamefont {SenGupta}},\ }\bibfield  {title} {\bibinfo {title} {Aspects of warped braneworld models},\ }\href@noop {} {\bibfield  {journal} {\bibinfo  {journal} {arXiv preprint arXiv:0812.1092}\ } (\bibinfo {year} {2008})}\BibitemShut {NoStop}%
\bibitem [{\citenamefont {Thiemann}(2003)}]{thiemann2003lectures}%
  \BibitemOpen
  \bibfield  {author} {\bibinfo {author} {\bibfnamefont {T.}~\bibnamefont {Thiemann}},\ }\bibfield  {title} {\bibinfo {title} {Lectures on loop quantum gravity},\ }in\ \href@noop {} {\emph {\bibinfo {booktitle} {Quantum gravity: From theory to experimental search}}}\ (\bibinfo  {publisher} {Springer},\ \bibinfo {year} {2003})\ pp.\ \bibinfo {pages} {41--135}\BibitemShut {NoStop}%
\bibitem [{\citenamefont {Mignemi}\ and\ \citenamefont {Stewart}(1993)}]{mignemi1993charged}%
  \BibitemOpen
  \bibfield  {author} {\bibinfo {author} {\bibfnamefont {S.}~\bibnamefont {Mignemi}}\ and\ \bibinfo {author} {\bibfnamefont {N.}~\bibnamefont {Stewart}},\ }\bibfield  {title} {\bibinfo {title} {Charged black holes in effective string theory},\ }\href@noop {} {\bibfield  {journal} {\bibinfo  {journal} {Physical Review D}\ }\textbf {\bibinfo {volume} {47}},\ \bibinfo {pages} {5259} (\bibinfo {year} {1993})}\BibitemShut {NoStop}%
\bibitem [{\citenamefont {Gibbons}\ and\ \citenamefont {Maeda}(1988)}]{gibbons1988black}%
  \BibitemOpen
  \bibfield  {author} {\bibinfo {author} {\bibfnamefont {G.~W.}\ \bibnamefont {Gibbons}}\ and\ \bibinfo {author} {\bibfnamefont {K.-i.}\ \bibnamefont {Maeda}},\ }\bibfield  {title} {\bibinfo {title} {Black holes and membranes in higher-dimensional theories with dilaton fields},\ }\href@noop {} {\bibfield  {journal} {\bibinfo  {journal} {Nuclear Physics B}\ }\textbf {\bibinfo {volume} {298}},\ \bibinfo {pages} {741} (\bibinfo {year} {1988})}\BibitemShut {NoStop}%
\bibitem [{\citenamefont {Sen}(1992)}]{sen1992rotating}%
  \BibitemOpen
  \bibfield  {author} {\bibinfo {author} {\bibfnamefont {A.}~\bibnamefont {Sen}},\ }\bibfield  {title} {\bibinfo {title} {Rotating charged black hole solution in heterotic string theory},\ }\href@noop {} {\bibfield  {journal} {\bibinfo  {journal} {Physical Review Letters}\ }\textbf {\bibinfo {volume} {69}},\ \bibinfo {pages} {1006} (\bibinfo {year} {1992})}\BibitemShut {NoStop}%
\bibitem [{\citenamefont {Bhadra}(2003)}]{Bhadra:2003zs}%
  \BibitemOpen
  \bibfield  {author} {\bibinfo {author} {\bibfnamefont {A.}~\bibnamefont {Bhadra}},\ }\bibfield  {title} {\bibinfo {title} {{Gravitational lensing by a charged black hole of string theory}},\ }\href {https://doi.org/10.1103/PhysRevD.67.103009} {\bibfield  {journal} {\bibinfo  {journal} {Phys. Rev. D}\ }\textbf {\bibinfo {volume} {67}},\ \bibinfo {pages} {103009} (\bibinfo {year} {2003})},\ \Eprint {https://arxiv.org/abs/gr-qc/0306016} {arXiv:gr-qc/0306016} \BibitemShut {NoStop}%
\bibitem [{\citenamefont {Gyulchev}\ and\ \citenamefont {Yazadjiev}(2007)}]{gyulchev2007kerr}%
  \BibitemOpen
  \bibfield  {author} {\bibinfo {author} {\bibfnamefont {G.~N.}\ \bibnamefont {Gyulchev}}\ and\ \bibinfo {author} {\bibfnamefont {S.~S.}\ \bibnamefont {Yazadjiev}},\ }\bibfield  {title} {\bibinfo {title} {Kerr-sen dilaton-axion black hole lensing in the strong deflection limit},\ }\href@noop {} {\bibfield  {journal} {\bibinfo  {journal} {Physical Review D—Particles, Fields, Gravitation, and Cosmology}\ }\textbf {\bibinfo {volume} {75}},\ \bibinfo {pages} {023006} (\bibinfo {year} {2007})}\BibitemShut {NoStop}%
\bibitem [{\citenamefont {Gyulchev}\ and\ \citenamefont {Yazadjiev}(2010)}]{gyulchev2010analytical}%
  \BibitemOpen
  \bibfield  {author} {\bibinfo {author} {\bibfnamefont {G.~N.}\ \bibnamefont {Gyulchev}}\ and\ \bibinfo {author} {\bibfnamefont {S.~S.}\ \bibnamefont {Yazadjiev}},\ }\bibfield  {title} {\bibinfo {title} {Analytical kerr-sen dilaton-axion black hole lensing in the weak deflection limit},\ }\href@noop {} {\bibfield  {journal} {\bibinfo  {journal} {Physical Review D—Particles, Fields, Gravitation, and Cosmology}\ }\textbf {\bibinfo {volume} {81}},\ \bibinfo {pages} {023005} (\bibinfo {year} {2010})}\BibitemShut {NoStop}%
\bibitem [{\citenamefont {Siahaan}(2016)}]{siahaan2016destroying}%
  \BibitemOpen
  \bibfield  {author} {\bibinfo {author} {\bibfnamefont {H.~M.}\ \bibnamefont {Siahaan}},\ }\bibfield  {title} {\bibinfo {title} {Destroying kerr-sen black holes},\ }\href@noop {} {\bibfield  {journal} {\bibinfo  {journal} {Physical Review D}\ }\textbf {\bibinfo {volume} {93}},\ \bibinfo {pages} {064028} (\bibinfo {year} {2016})}\BibitemShut {NoStop}%
\bibitem [{\citenamefont {Wald}(1974)}]{wald1974gedanken}%
  \BibitemOpen
  \bibfield  {author} {\bibinfo {author} {\bibfnamefont {R.}~\bibnamefont {Wald}},\ }\bibfield  {title} {\bibinfo {title} {Gedanken experiments to destroy a black hole},\ }\href@noop {} {\bibfield  {journal} {\bibinfo  {journal} {Annals of Physics}\ }\textbf {\bibinfo {volume} {82}},\ \bibinfo {pages} {548} (\bibinfo {year} {1974})}\BibitemShut {NoStop}%
\bibitem [{\citenamefont {Xavier}\ \emph {et~al.}(2020)\citenamefont {Xavier}, \citenamefont {Cunha}, \citenamefont {Crispino},\ and\ \citenamefont {Herdeiro}}]{xavier2020shadows}%
  \BibitemOpen
  \bibfield  {author} {\bibinfo {author} {\bibfnamefont {S.~V. M.~C.}\ \bibnamefont {Xavier}}, \bibinfo {author} {\bibfnamefont {P.~V.}\ \bibnamefont {Cunha}}, \bibinfo {author} {\bibfnamefont {L.~C.}\ \bibnamefont {Crispino}},\ and\ \bibinfo {author} {\bibfnamefont {C.~A.}\ \bibnamefont {Herdeiro}},\ }\bibfield  {title} {\bibinfo {title} {Shadows of charged rotating black holes: Kerr--newman versus kerr--sen},\ }\href@noop {} {\bibfield  {journal} {\bibinfo  {journal} {International Journal of Modern Physics D}\ }\textbf {\bibinfo {volume} {29}},\ \bibinfo {pages} {2041005} (\bibinfo {year} {2020})}\BibitemShut {NoStop}%
\bibitem [{\citenamefont {Guo}\ \emph {et~al.}(2020)\citenamefont {Guo}, \citenamefont {Song},\ and\ \citenamefont {Yan}}]{guo2020observational}%
  \BibitemOpen
  \bibfield  {author} {\bibinfo {author} {\bibfnamefont {M.}~\bibnamefont {Guo}}, \bibinfo {author} {\bibfnamefont {S.}~\bibnamefont {Song}},\ and\ \bibinfo {author} {\bibfnamefont {H.}~\bibnamefont {Yan}},\ }\bibfield  {title} {\bibinfo {title} {Observational signature of a near-extremal kerr-sen black hole in the heterotic string theory},\ }\href@noop {} {\bibfield  {journal} {\bibinfo  {journal} {Physical Review D}\ }\textbf {\bibinfo {volume} {101}},\ \bibinfo {pages} {024055} (\bibinfo {year} {2020})}\BibitemShut {NoStop}%
\bibitem [{\citenamefont {Uniyal}\ \emph {et~al.}(2017)\citenamefont {Uniyal}, \citenamefont {Nandan},\ and\ \citenamefont {Purohit}}]{uniyal2017null}%
  \BibitemOpen
  \bibfield  {author} {\bibinfo {author} {\bibfnamefont {R.}~\bibnamefont {Uniyal}}, \bibinfo {author} {\bibfnamefont {H.}~\bibnamefont {Nandan}},\ and\ \bibinfo {author} {\bibfnamefont {K.}~\bibnamefont {Purohit}},\ }\bibfield  {title} {\bibinfo {title} {Null geodesics and observables around the kerr--sen black hole},\ }\href@noop {} {\bibfield  {journal} {\bibinfo  {journal} {Classical and Quantum Gravity}\ }\textbf {\bibinfo {volume} {35}},\ \bibinfo {pages} {025003} (\bibinfo {year} {2017})}\BibitemShut {NoStop}%
\bibitem [{\citenamefont {Uniyal}\ \emph {et~al.}(2018)\citenamefont {Uniyal}, \citenamefont {Nandan},\ and\ \citenamefont {Jetzer}}]{uniyal2018bending}%
  \BibitemOpen
  \bibfield  {author} {\bibinfo {author} {\bibfnamefont {R.}~\bibnamefont {Uniyal}}, \bibinfo {author} {\bibfnamefont {H.}~\bibnamefont {Nandan}},\ and\ \bibinfo {author} {\bibfnamefont {P.}~\bibnamefont {Jetzer}},\ }\bibfield  {title} {\bibinfo {title} {Bending angle of light in equatorial plane of kerr--sen black hole},\ }\href@noop {} {\bibfield  {journal} {\bibinfo  {journal} {Physics Letters B}\ }\textbf {\bibinfo {volume} {782}},\ \bibinfo {pages} {185} (\bibinfo {year} {2018})}\BibitemShut {NoStop}%
\bibitem [{\citenamefont {Misner}\ \emph {et~al.}(1973)\citenamefont {Misner}, \citenamefont {Thorne},\ and\ \citenamefont {Wheeler}}]{misner1973gravitation}%
  \BibitemOpen
  \bibfield  {author} {\bibinfo {author} {\bibfnamefont {C.~W.}\ \bibnamefont {Misner}}, \bibinfo {author} {\bibfnamefont {K.~S.}\ \bibnamefont {Thorne}},\ and\ \bibinfo {author} {\bibfnamefont {J.~A.}\ \bibnamefont {Wheeler}},\ }\href@noop {} {\emph {\bibinfo {title} {Gravitation}}}\ (\bibinfo  {publisher} {Macmillan},\ \bibinfo {year} {1973})\BibitemShut {NoStop}%
\bibitem [{\citenamefont {Weinberg}(1972)}]{weinberg1972principles}%
  \BibitemOpen
  \bibfield  {author} {\bibinfo {author} {\bibfnamefont {S.}~\bibnamefont {Weinberg}},\ }\href@noop {} {\emph {\bibinfo {title} {Principles and applications of the general theory of relativity: gravitation and cosmology}}}\ (\bibinfo  {publisher} {Wiley},\ \bibinfo {year} {1972})\BibitemShut {NoStop}%
\bibitem [{\citenamefont {Schneider}\ \emph {et~al.}(1992)\citenamefont {Schneider}, \citenamefont {Ehlers},\ and\ \citenamefont {Falco}}]{schneider1992gravitational}%
  \BibitemOpen
  \bibfield  {author} {\bibinfo {author} {\bibfnamefont {P.}~\bibnamefont {Schneider}}, \bibinfo {author} {\bibfnamefont {J.}~\bibnamefont {Ehlers}},\ and\ \bibinfo {author} {\bibfnamefont {E.}~\bibnamefont {Falco}},\ }\bibfield  {title} {\bibinfo {title} {Gravitational lenses springer-verlag},\ }\href@noop {} {\bibfield  {journal} {\bibinfo  {journal} {Berlin Inc., New York}\ } (\bibinfo {year} {1992})}\BibitemShut {NoStop}%
\bibitem [{\citenamefont {Tamm}(1924)}]{tamm1924electrodynamics}%
  \BibitemOpen
  \bibfield  {author} {\bibinfo {author} {\bibfnamefont {I.}~\bibnamefont {Tamm}},\ }\bibfield  {title} {\bibinfo {title} {The electrodynamics of anisotropic media in the special theory of relativity},\ }\href@noop {} {\bibfield  {journal} {\bibinfo  {journal} {Zh. Rus. Fiz.-Khim. Obshchestva, Otd. Fiz}\ }\textbf {\bibinfo {volume} {56}},\ \bibinfo {pages} {248} (\bibinfo {year} {1924})}\BibitemShut {NoStop}%
\bibitem [{\citenamefont {Balazs}(1958)}]{balazs1958effect}%
  \BibitemOpen
  \bibfield  {author} {\bibinfo {author} {\bibfnamefont {N.}~\bibnamefont {Balazs}},\ }\bibfield  {title} {\bibinfo {title} {Effect of a gravitational field, due to a rotating body, on the plane of polarization of an electromagnetic wave},\ }\href@noop {} {\bibfield  {journal} {\bibinfo  {journal} {Physical Review}\ }\textbf {\bibinfo {volume} {110}},\ \bibinfo {pages} {236} (\bibinfo {year} {1958})}\BibitemShut {NoStop}%
\bibitem [{\citenamefont {Plebanski}(1960)}]{plebanski1960electromagnetic}%
  \BibitemOpen
  \bibfield  {author} {\bibinfo {author} {\bibfnamefont {J.}~\bibnamefont {Plebanski}},\ }\bibfield  {title} {\bibinfo {title} {Electromagnetic waves in gravitational fields},\ }\href@noop {} {\bibfield  {journal} {\bibinfo  {journal} {Physical Review}\ }\textbf {\bibinfo {volume} {118}},\ \bibinfo {pages} {1396} (\bibinfo {year} {1960})}\BibitemShut {NoStop}%
\bibitem [{\citenamefont {Mashhoon}(1975)}]{mashhoon1975influence}%
  \BibitemOpen
  \bibfield  {author} {\bibinfo {author} {\bibfnamefont {B.}~\bibnamefont {Mashhoon}},\ }\bibfield  {title} {\bibinfo {title} {Influence of gravitation on the propagation of electromagnetic radiation},\ }\href@noop {} {\bibfield  {journal} {\bibinfo  {journal} {Physical Review D}\ }\textbf {\bibinfo {volume} {11}},\ \bibinfo {pages} {2679} (\bibinfo {year} {1975})}\BibitemShut {NoStop}%
\bibitem [{\citenamefont {Fischbach}\ and\ \citenamefont {Freeman}(1980)}]{fischbach1980second}%
  \BibitemOpen
  \bibfield  {author} {\bibinfo {author} {\bibfnamefont {E.}~\bibnamefont {Fischbach}}\ and\ \bibinfo {author} {\bibfnamefont {B.~S.}\ \bibnamefont {Freeman}},\ }\bibfield  {title} {\bibinfo {title} {Second-order contribution to the gravitational deflection of light},\ }\href@noop {} {\bibfield  {journal} {\bibinfo  {journal} {Physical Review D}\ }\textbf {\bibinfo {volume} {22}},\ \bibinfo {pages} {2950} (\bibinfo {year} {1980})}\BibitemShut {NoStop}%
\bibitem [{\citenamefont {Evans}\ and\ \citenamefont {Rosenquist}(1986)}]{evans1986f}%
  \BibitemOpen
  \bibfield  {author} {\bibinfo {author} {\bibfnamefont {J.}~\bibnamefont {Evans}}\ and\ \bibinfo {author} {\bibfnamefont {M.}~\bibnamefont {Rosenquist}},\ }\bibfield  {title} {\bibinfo {title} {F= ma optics},\ }\href@noop {} {\bibfield  {journal} {\bibinfo  {journal} {Am. J. Phys}\ }\textbf {\bibinfo {volume} {54}},\ \bibinfo {pages} {876} (\bibinfo {year} {1986})}\BibitemShut {NoStop}%
\bibitem [{\citenamefont {Evans}\ \emph {et~al.}(1996)\citenamefont {Evans}, \citenamefont {Nandi},\ and\ \citenamefont {Islam}}]{evans1996optical}%
  \BibitemOpen
  \bibfield  {author} {\bibinfo {author} {\bibfnamefont {J.}~\bibnamefont {Evans}}, \bibinfo {author} {\bibfnamefont {K.~K.}\ \bibnamefont {Nandi}},\ and\ \bibinfo {author} {\bibfnamefont {A.}~\bibnamefont {Islam}},\ }\bibfield  {title} {\bibinfo {title} {The optical--mechanical analogy in general relativity: New methods for the paths of light and of the planets},\ }\href@noop {} {\bibfield  {journal} {\bibinfo  {journal} {American Journal of Physics}\ }\textbf {\bibinfo {volume} {64}},\ \bibinfo {pages} {1404} (\bibinfo {year} {1996})}\BibitemShut {NoStop}%
\bibitem [{\citenamefont {Sereno}(2003)}]{sereno2003gravitational}%
  \BibitemOpen
  \bibfield  {author} {\bibinfo {author} {\bibfnamefont {M.}~\bibnamefont {Sereno}},\ }\bibfield  {title} {\bibinfo {title} {Gravitational lensing in metric theories of gravity},\ }\href@noop {} {\bibfield  {journal} {\bibinfo  {journal} {Physical Review D}\ }\textbf {\bibinfo {volume} {67}},\ \bibinfo {pages} {064007} (\bibinfo {year} {2003})}\BibitemShut {NoStop}%
\bibitem [{\citenamefont {Sereno}(2004)}]{sereno2004weak}%
  \BibitemOpen
  \bibfield  {author} {\bibinfo {author} {\bibfnamefont {M.}~\bibnamefont {Sereno}},\ }\bibfield  {title} {\bibinfo {title} {Weak field limit of reissner-nordstr{\"o}m black hole lensing},\ }\href@noop {} {\bibfield  {journal} {\bibinfo  {journal} {Physical Review D}\ }\textbf {\bibinfo {volume} {69}},\ \bibinfo {pages} {023002} (\bibinfo {year} {2004})}\BibitemShut {NoStop}%
\bibitem [{\citenamefont {Sen}(2010)}]{sen2010more}%
  \BibitemOpen
  \bibfield  {author} {\bibinfo {author} {\bibfnamefont {A.}~\bibnamefont {Sen}},\ }\bibfield  {title} {\bibinfo {title} {A more exact expression for the gravitational deflection of light, derived using material medium approach},\ }\href@noop {} {\bibfield  {journal} {\bibinfo  {journal} {Astrophysics}\ }\textbf {\bibinfo {volume} {53}},\ \bibinfo {pages} {560} (\bibinfo {year} {2010})}\BibitemShut {NoStop}%
\bibitem [{\citenamefont {Roy}\ and\ \citenamefont {Sen}(2015)}]{roy2015trajectory}%
  \BibitemOpen
  \bibfield  {author} {\bibinfo {author} {\bibfnamefont {S.}~\bibnamefont {Roy}}\ and\ \bibinfo {author} {\bibfnamefont {A.}~\bibnamefont {Sen}},\ }\bibfield  {title} {\bibinfo {title} {Trajectory of a light ray in kerr field: A material medium approach},\ }\href@noop {} {\bibfield  {journal} {\bibinfo  {journal} {Astrophysics and Space Science}\ }\textbf {\bibinfo {volume} {360}},\ \bibinfo {pages} {1} (\bibinfo {year} {2015})}\BibitemShut {NoStop}%
\bibitem [{\citenamefont {Roy}\ and\ \citenamefont {Sen}(2017)}]{roy2017deflection}%
  \BibitemOpen
  \bibfield  {author} {\bibinfo {author} {\bibfnamefont {S.}~\bibnamefont {Roy}}\ and\ \bibinfo {author} {\bibfnamefont {A.}~\bibnamefont {Sen}},\ }\bibfield  {title} {\bibinfo {title} {Deflection of light ray due to a charged body using material medium approach},\ }\href@noop {} {\bibfield  {journal} {\bibinfo  {journal} {Zeitschrift f{\"u}r Naturforschung A}\ }\textbf {\bibinfo {volume} {72}},\ \bibinfo {pages} {1113} (\bibinfo {year} {2017})}\BibitemShut {NoStop}%
\bibitem [{\citenamefont {del Barco}(2024)}]{del2024accurate}%
  \BibitemOpen
  \bibfield  {author} {\bibinfo {author} {\bibfnamefont {O.}~\bibnamefont {del Barco}},\ }\bibfield  {title} {\bibinfo {title} {An accurate equation for the gravitational bending of light by a static massive object},\ }\href@noop {} {\bibfield  {journal} {\bibinfo  {journal} {Monthly Notices of the Royal Astronomical Society}\ }\textbf {\bibinfo {volume} {535}},\ \bibinfo {pages} {2504} (\bibinfo {year} {2024})}\BibitemShut {NoStop}%
\bibitem [{\citenamefont {Ruggiero}(2025)}]{ruggiero2025effects}%
  \BibitemOpen
  \bibfield  {author} {\bibinfo {author} {\bibfnamefont {M.~L.}\ \bibnamefont {Ruggiero}},\ }\bibfield  {title} {\bibinfo {title} {Effects of gravitational waves on electromagnetic fields},\ }\href@noop {} {\bibfield  {journal} {\bibinfo  {journal} {The European Physical Journal C}\ }\textbf {\bibinfo {volume} {85}},\ \bibinfo {pages} {1} (\bibinfo {year} {2025})}\BibitemShut {NoStop}%
\bibitem [{\citenamefont {Wu}\ \emph {et~al.}(2020)\citenamefont {Wu}, \citenamefont {Wu}, \citenamefont {Yu},\ and\ \citenamefont {Wu}}]{wu2020ultraspinning}%
  \BibitemOpen
  \bibfield  {author} {\bibinfo {author} {\bibfnamefont {D.}~\bibnamefont {Wu}}, \bibinfo {author} {\bibfnamefont {P.}~\bibnamefont {Wu}}, \bibinfo {author} {\bibfnamefont {H.}~\bibnamefont {Yu}},\ and\ \bibinfo {author} {\bibfnamefont {S.-Q.}\ \bibnamefont {Wu}},\ }\bibfield  {title} {\bibinfo {title} {Are ultraspinning kerr-sen-ads 4 black holes always superentropic?},\ }\href@noop {} {\bibfield  {journal} {\bibinfo  {journal} {Physical Review D}\ }\textbf {\bibinfo {volume} {102}},\ \bibinfo {pages} {044007} (\bibinfo {year} {2020})}\BibitemShut {NoStop}%
\bibitem [{\citenamefont {Burinskii}(1995)}]{burinskii1995some}%
  \BibitemOpen
  \bibfield  {author} {\bibinfo {author} {\bibfnamefont {A.~Y.}\ \bibnamefont {Burinskii}},\ }\bibfield  {title} {\bibinfo {title} {Some properties of the kerr solution to low energy string theory},\ }\href@noop {} {\bibfield  {journal} {\bibinfo  {journal} {Physical Review D}\ }\textbf {\bibinfo {volume} {52}},\ \bibinfo {pages} {5826} (\bibinfo {year} {1995})}\BibitemShut {NoStop}%
\bibitem [{\citenamefont {Griffiths}\ and\ \citenamefont {Podolsk{\`y}}(2009)}]{griffiths2009exact}%
  \BibitemOpen
  \bibfield  {author} {\bibinfo {author} {\bibfnamefont {J.~B.}\ \bibnamefont {Griffiths}}\ and\ \bibinfo {author} {\bibfnamefont {J.}~\bibnamefont {Podolsk{\`y}}},\ }\href@noop {} {\emph {\bibinfo {title} {Exact space-times in Einstein's general relativity}}}\ (\bibinfo  {publisher} {Cambridge University Press},\ \bibinfo {year} {2009})\BibitemShut {NoStop}%
\bibitem [{\citenamefont {Zhang}\ and\ \citenamefont {Jiang}(2021)}]{zhang2021escape}%
  \BibitemOpen
  \bibfield  {author} {\bibinfo {author} {\bibfnamefont {M.}~\bibnamefont {Zhang}}\ and\ \bibinfo {author} {\bibfnamefont {J.}~\bibnamefont {Jiang}},\ }\bibfield  {title} {\bibinfo {title} {Escape probability of particle from kerr-sen black hole},\ }\href@noop {} {\bibfield  {journal} {\bibinfo  {journal} {Nuclear Physics B}\ }\textbf {\bibinfo {volume} {964}},\ \bibinfo {pages} {115313} (\bibinfo {year} {2021})}\BibitemShut {NoStop}%
\bibitem [{\citenamefont {Feng}\ and\ \citenamefont {L{\"u}}(2020)}]{feng2020size}%
  \BibitemOpen
  \bibfield  {author} {\bibinfo {author} {\bibfnamefont {X.-H.}\ \bibnamefont {Feng}}\ and\ \bibinfo {author} {\bibfnamefont {H.}~\bibnamefont {L{\"u}}},\ }\bibfield  {title} {\bibinfo {title} {On the size of rotating black holes},\ }\href@noop {} {\bibfield  {journal} {\bibinfo  {journal} {The European Physical Journal C}\ }\textbf {\bibinfo {volume} {80}},\ \bibinfo {pages} {1} (\bibinfo {year} {2020})}\BibitemShut {NoStop}%
\bibitem [{\citenamefont {Landau}(2013)}]{landau2013classical}%
  \BibitemOpen
  \bibfield  {author} {\bibinfo {author} {\bibfnamefont {L.~D.}\ \bibnamefont {Landau}},\ }\href@noop {} {\emph {\bibinfo {title} {The classical theory of fields}}},\ Vol.~\bibinfo {volume} {2}\ (\bibinfo  {publisher} {Elsevier},\ \bibinfo {year} {2013})\BibitemShut {NoStop}%
\bibitem [{\citenamefont {Born}\ and\ \citenamefont {Wolf}(2013)}]{born2013principles}%
  \BibitemOpen
  \bibfield  {author} {\bibinfo {author} {\bibfnamefont {M.}~\bibnamefont {Born}}\ and\ \bibinfo {author} {\bibfnamefont {E.}~\bibnamefont {Wolf}},\ }\href@noop {} {\emph {\bibinfo {title} {Principles of optics: electromagnetic theory of propagation, interference and diffraction of light}}}\ (\bibinfo  {publisher} {Elsevier},\ \bibinfo {year} {2013})\BibitemShut {NoStop}%
\bibitem [{\citenamefont {Chandrasekhar}(1998)}]{Chandrasekhar1998}%
  \BibitemOpen
  \bibfield  {author} {\bibinfo {author} {\bibfnamefont {S.}~\bibnamefont {Chandrasekhar}},\ }\href@noop {} {\emph {\bibinfo {title} {The mathematical theory of black holes}}},\ Vol.~\bibinfo {volume} {69}\ (\bibinfo  {publisher} {Oxford University Press},\ \bibinfo {year} {1998})\BibitemShut {NoStop}%
\bibitem [{\citenamefont {Claudel}\ \emph {et~al.}(2001)\citenamefont {Claudel}, \citenamefont {Virbhadra},\ and\ \citenamefont {Ellis}}]{Claudel:2000yi}%
  \BibitemOpen
  \bibfield  {author} {\bibinfo {author} {\bibfnamefont {C.-M.}\ \bibnamefont {Claudel}}, \bibinfo {author} {\bibfnamefont {K.~S.}\ \bibnamefont {Virbhadra}},\ and\ \bibinfo {author} {\bibfnamefont {G.~F.~R.}\ \bibnamefont {Ellis}},\ }\bibfield  {title} {\bibinfo {title} {{The Geometry of photon surfaces}},\ }\href {https://doi.org/10.1063/1.1308507} {\bibfield  {journal} {\bibinfo  {journal} {J. Math. Phys.}\ }\textbf {\bibinfo {volume} {42}},\ \bibinfo {pages} {818} (\bibinfo {year} {2001})},\ \Eprint {https://arxiv.org/abs/gr-qc/0005050} {arXiv:gr-qc/0005050} \BibitemShut {NoStop}%
\bibitem [{\citenamefont {Iyer}\ and\ \citenamefont {Hansen}(2009)}]{Iyer:2009wa}%
  \BibitemOpen
  \bibfield  {author} {\bibinfo {author} {\bibfnamefont {S.~V.}\ \bibnamefont {Iyer}}\ and\ \bibinfo {author} {\bibfnamefont {E.~C.}\ \bibnamefont {Hansen}},\ }\bibfield  {title} {\bibinfo {title} {{Light's Bending Angle in the Equatorial Plane of a Kerr Black Hole}},\ }\href {https://doi.org/10.1103/PhysRevD.80.124023} {\bibfield  {journal} {\bibinfo  {journal} {Phys. Rev. D}\ }\textbf {\bibinfo {volume} {80}},\ \bibinfo {pages} {124023} (\bibinfo {year} {2009})},\ \Eprint {https://arxiv.org/abs/0907.5352} {arXiv:0907.5352 [gr-qc]} \BibitemShut {NoStop}%
\bibitem [{\citenamefont {Hsiao}\ \emph {et~al.}(2020)\citenamefont {Hsiao}, \citenamefont {Lee},\ and\ \citenamefont {Lin}}]{Hsiao:2019ohy}%
  \BibitemOpen
  \bibfield  {author} {\bibinfo {author} {\bibfnamefont {Y.-W.}\ \bibnamefont {Hsiao}}, \bibinfo {author} {\bibfnamefont {D.-S.}\ \bibnamefont {Lee}},\ and\ \bibinfo {author} {\bibfnamefont {C.-Y.}\ \bibnamefont {Lin}},\ }\bibfield  {title} {\bibinfo {title} {{Equatorial light bending around Kerr-Newman black holes}},\ }\href {https://doi.org/10.1103/PhysRevD.101.064070} {\bibfield  {journal} {\bibinfo  {journal} {Phys. Rev. D}\ }\textbf {\bibinfo {volume} {101}},\ \bibinfo {pages} {064070} (\bibinfo {year} {2020})},\ \Eprint {https://arxiv.org/abs/1910.04372} {arXiv:1910.04372 [gr-qc]} \BibitemShut {NoStop}%
\end{thebibliography}%

\end{document}